\documentclass[AMA,STIX1COL,12pt]{WileyNJD-v2}
\usepackage{moreverb}
\usepackage{mathtools}
\usepackage{float}

 \newcommand{\ind}{\perp\!\!\!\!\perp} 

\newcommand\BibTeX{{\rmfamily B\kern-.05em \textsc{i\kern-.025em b}\kern-.08em
T\kern-.1667em\lower.7ex\hbox{E}\kern-.125emX}}

\articletype{Research Article}%

\received{<day> <Month>, <year>}
\revised{<day> <Month>, <year>}
\accepted{<day> <Month>, <year>}

\begin{document}

\title{Causal Mediation Analysis for Zero-inflated Mixture Mediators}

\author[1]{Meilin Jiang*}
\author[2,3]{Seonjoo Lee}
\author[4,5]{A. James O'Malley}
\author[6]{Pengfei Li}
\author[7]{Zhigang Li}

\authormark{Jiang \textsc{et al}}

\address[1]{\orgname{Eli Lilly and Company}, \orgaddress{\state{Indianapolis, IN}, \country{USA}}}
\address[2]{\orgdiv{Mental Health Data Science}, \orgname{New York State Psychiatric Institute}, \orgaddress{\state{New York, NY}, \country{USA}}}
\address[3]{\orgdiv{Departments of Biostatistics and Psychiatry}, \orgname{Columbia University}, \orgaddress{\state{New York, NY}, \country{USA}}}
\address[4]{\orgdiv{Department of Biomedical Data Science}, \orgname{Geisel School of Medicine at Dartmouth}, \orgaddress{\state{Hanover, NH}, \country{USA}}}
\address[5]{\orgdiv{The Dartmouth Institute}, \orgname{Geisel School of Medicine at Dartmouth}, \orgaddress{\state{Hanover, NH}, \country{USA}}}
\address[6]{\orgdiv{Department of Statistics}, \orgname{University of Waterloo}, \orgaddress{\state{Waterloo, Ontario}, \country{Canada}}}
\address[7]{\orgdiv{Department of Biostatistics}, \orgname{University of Florida}, \orgaddress{\state{Gainesville, FL}, \country{USA}}}

\corres{*Meilin Jiang, Eli Lilly and Company, Indianapolis, IN, USA.  \email{meilin.jiang@lilly.com}}

\presentaddress{Eli Lilly and Company, Indianapolis, IN, USA}

\abstract[Abstract]{
Causal mediation analysis is an important statistical tool to quantify effects transmitted by intermediate variables from a cause to an outcome. There is a gap in mediation analysis methods to handle mixture mediator data that are zero-inflated with multi-modality and atypical behaviors. We propose an innovative way to model zero-inflated mixture mediators from the perspective of finite mixture distributions to flexibly capture such mediator data. Multiple data types are considered for modeling such mediators including the zero-inflated log-normal mixture, zero-inflated Poisson mixture and zero-inflated negative binomial mixture. A two-part mediation effect is derived to better understand effects on outcomes attributable to the numerical change as well as binary change from 0 to 1 in mediators. The maximum likelihood estimates are obtained by an expectation–maximization algorithm to account for unobserved mixture membership and whether an observed zero is a true or false zero. The optimal number of mixture components are chosen by a model selection criterion. The performance of the proposed method is demonstrated in a simulation study and an application to a neuroscience study in comparison with standard mediation analysis methods.
}



\maketitle


\section{Introduction}
Causal mediation analysis is a useful statistical tool to estimate the amount of effect transmitted by a mediator variable from a cause to its outcome. In addition to a deeper understanding of a causal mechanism, another important motivation to study mediation effects (i.e., indirect effects) is to bring benefits from the perspective of mediators. In some cases, the modification of a cause or independent variable is not feasible or ethical, but the effect on the outcome from a mediator can be targeted instead, which could be potentially more efficient \citep{Vanderweele2015}. Mediation analysis started to gain popularity with the classical ``product of coefficients method" using linear regression approaches \citep{Baron1986,Sobel1982}. Later, the counterfactual/potential outcomes approach \citep{Robins1992,Pearl2001,VanderWeele2009,Imai2010} became a main focus of mediation analyses given its suitability for non-linear models with possible interactions between independent variables and mediators. With desirable data types and assumed distributions, the mediation effect can have a closed-form expression \citep{VanderWeele2009}, while simulation based inference is also available for cases with less distributional assumptions \citep{Imai2010}.

Given the increasing availability of various data types, mediation analysis models to handle mixture mediators with special behaviors are of great interest, especially for mixture mediator data coming from uncommon distributions with zero-inflation, multi-modality, and atypical behaviors. In practice, mixture data are very common because data might be sampled from more than one underlying population. For example, blood pressure measurements will be a mixture when the sample consists of patients with different stages of hypertension. Moreover, mediator variables measured with excessive zeros are commonly encountered in biomedical studies such as neuroscience data \citep{Li2021,moura2019relationship,Hellton2020}, and there are existing approaches to deal with zero-inflated mediators \citep{wuMediation,Jiang2023}. However, as shown in our motivation example, the zero-inflated neuroimage data also exhibit mixture features and multi-modality. There are a lack of approaches to model such data.
As a classical statistical problem, mixture distributions have been studied in terms of their properties and estimation since the 19th century in order to separate a population into more homogeneous subgroups \citep{FISHER1936,N.E.1969,Zhang2013}. There are two types of mixture distribution: infinite mixtures in which some distributions are averaged over a continuous mixing distribution and finite mixtures with a discrete number of mixing components. The finite mixture distribution is widely used to model data with unusual shapes that cannot be approximated by common distributions. Compared to a single distribution, a mixture distribution is more flexible to capture data with skewness, multi-modality, and atypical behaviors \citep{McLachlan2019}. It relaxes limits on typically-used statistical models. For instance, the Gaussian mixture distribution is one of the most frequently used mixture distributions given its flexibility. Multiple methods exist to estimate parameters from a mixture distribution such as method of moments, moment generating functions, and characteristic functions \citep{N.E.1969,Quandt1978}. Among them, the Expectation-Maximization (EM) algorithm is the traditional approach to compute maximum likelihood estimates (MLE) given its desirable properties and guaranteed convergence with good initial values \citep{Redner1984}. 

This paper proposes an innovative way to handle zero-inflated mixture mediators. More specifically, zero-inflated log-normal (ZILoNM), zero-inflated Poisson (ZIPM), and zero-inflated negative binomial (ZINBM) mixtures are considered as options for modeling such mediators. The mediation effect can be decomposed into two parts due to its zero-inflated nature. The MLEs of model parameters are obtained from an EM algorithm to solve the following two challenges in estimation: Firstly, for a non-zero observation, we do not know which component of the mixture distribution it comes from; Secondly, we do not know if an observed zero value is a true or false zero since the excessive zeros in the data can be composed of both. A true zero implies that the measurement is indeed zero while a false zero means the underlying value is non-zero but is observed as a zero incorrectly. The chance of observing false zeros is accounted by specifying a probability mechanism. The Bayesian information criterion (BIC) is employed to select the optimal number of mixture components.

This paper is organized as follows. We begin by introducing a real world study that motivates the proposed method in section \ref{sc:moti}. Section \ref{sc:effects} introduces definitions of direct and indirect effects under the counterfactual outcomes framework. We present the proposed model specifications and required assumptions in Section \ref{sc:model}. Section \ref{sc:est} describes the estimation approach. A simulation study with results is shown in section \ref{sc:simu}. Section \ref{sc:application} provides a real data application. The discussion is included in section \ref{sc:diss} followed by the Appendix in section \ref{sc:appendix}.

\section{Motivating example - ABCD study}\label{sc:moti}
We introduce a real world example in neuroscience involving zero-inflated mixture mediators. Adolescent Brain Cognitive Development (ABCD) is a longitudinal research study (\url{https://abcdstudy.org}) on a large youth sample designed to study the brain development and child health in the United States \citep{Karcher2020}. 

Literature suggests that child behavior problems predict cognitive ability in later years and vice versa \citep{Zhang2023}.
In this data, we would like to examine whether brain structural connectivities mediate the effect of children's behavior problems on later age cognitive development, assessed using NIH Toolbox. Children's behavior problems were assessed using Child Behavior Checklist (CBCL) in depression, anxiety disorder, somatic problems, attention-deficit/hyperactivity disorder, oppositional defiant disorder, conduct problems, and obsessive-compulsive problems. There has been some research indicating a negative association between children's behavior problems and academic achievement \citep{Kremer2016}. 
The potential mediator, brain structural connectivities, were measured between different brain regions. The diffusion weighted imaging data were processed using MRtrix3 \citep{Tournier2019}. Briefly, preprocessing steps include eddy current correction, motion correction, $B_0$ distortion correction, gradient nonlinearities distortion correction. For tractography, we used the 2nd order integration over fiber orientation distributions algorithm the anatomically constrained tractography framework \citep{Smith2012}. After whole brain tractography generation, we obtained connectome matrices by mapping the streamlines based on their assignments to the node wise endpoints in the Glasser atlas \citep{Glasser2016} for cortical regions and aseg atlas \citep{Fischl2002} for subcortical regions. The cortical nodes were further aggregated into networks defined as Cole's networks \citep{Ji2019}. For illustration purpose, we considered two mediators: connectivity between right language network and left accumbens area and between left auditory network and right cerebellum.

\begin{figure}
  \begin{center}
  \includegraphics[width =1\textwidth,angle=0]{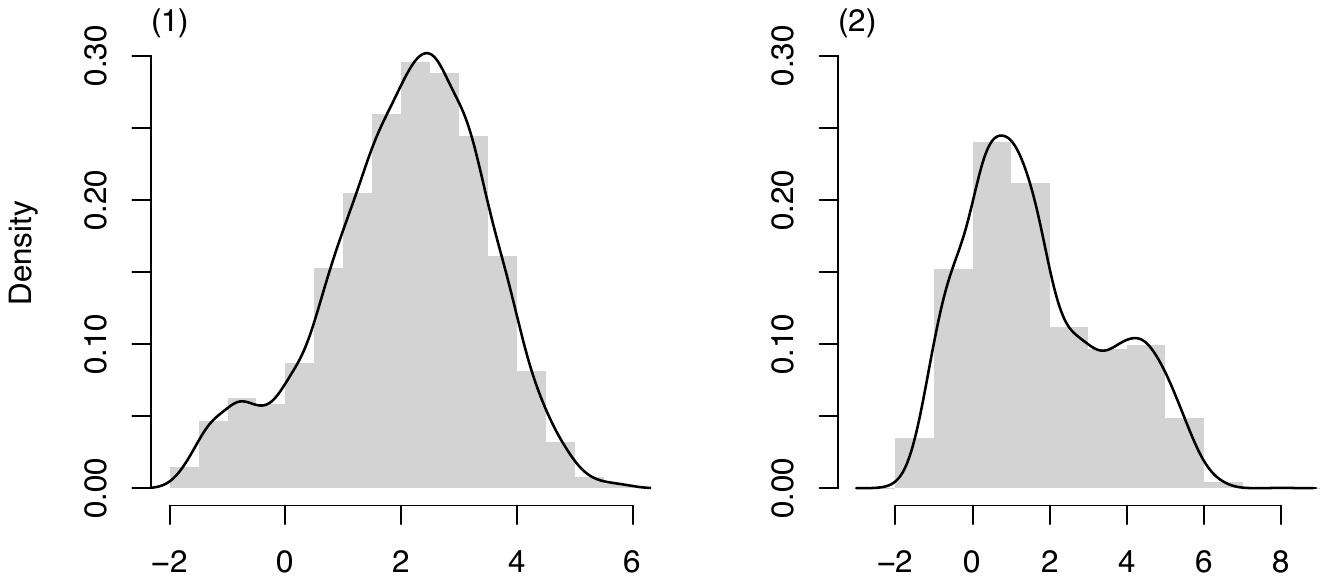}
  \end{center}
  \caption{Distribution of log transformed non-zero brain structural connectivities values (1) between right language network and left accumbens area (2) between left auditory network and right cerebellum in ABCD study.}
  \label{fig_ABCD}
\end{figure}

Nevertheless, difficulties arise from the atypical distribution of mediator data (i.e., brain connectivity). Firstly, zero-inflated values are commonly seen in brain connectivity as it could be rare to develop high connectivity between all brain regions. The two mediators of interest are composed of 29\% to 34\% zero values respectively in the sample of 8749 children. Moreover, both true and false zeros can be present leading to excessive zero values. True zeros come from no connectivities between brain regions at the time of assessment, while false zeros come from connectivity values that are too low to be detected. Secondly, the non-zero brain connectivity values are highly skewed on the original scale and may be generated from more than one distribution. As shown in Figure \ref{fig_ABCD}, we can observe that the distribution of log-non-zero values of the mediators shows two modes indicating a mixture data distribution. We propose a new mediation modeling approach to accommodate such mediators.

\section{Definitions of direct and indirect effects}\label{sc:effects}
Consider an independent variable $X$, a zero-inflated mediator $M$, and an outcome variable $Y$. 
Based on the zero-inflated structure of M, it can be written as the product of two random variables as follows:
\begin{align}
&M=B \cdot M^p,  \label{defM}
\end{align}
where $B$ takes value of 0 or 1 and follows a Bernoulli distribution, $M^p$ is the positive part of M, and $B$ and $M^p$ can be dependent. When $B=0$, the true value of $M$ is truly 0 as well (i.e., $B$ doesn't mask the true value of $M$). The false zeros will be generated by the probability mechanism given in subsection \ref{pfalse0}. We assume that $B$ is the cause of the zero-inflated structure for $M$. Notice that this is different than \cite{Jiang2023} where $M$ is the cause of the indicator variable $1_{(M>0)}$.
Let $Y_{xbm}$ denote the potential outcome of $Y$ when $(X, B, M)$ take the value $(x, b, m)$. Let $B_x$ and $M_x$ represent the value of $B$ and $M$, respectively, if $X$ takes the value $x$. Since $B$ and $M$ can be considered two sequential mediators, we write $M_x$ as $M(x,B_x)$.

Under the counterfactual outcomes framework \citep{Pearl2001,VanderWeele2009,Imai2010}, the average natural indirect effect (NIE) and natural direct effects (NDE) for $X$ changing from $x_1$ to $x_2$ are defined as:
\begin{align}
&\text{NIE}=E\big(Y_{x_2B_{x_2}M(x_2,B_{x_2})}-Y_{x_2B_{x_1}M(x_1,B_{x_1})}\big), \label{defNIE}\\
&\text{NDE}=E\big(Y_{x_2 B_{x_1}M(x_1,B_{x_1})}-Y_{x_1 B_{x_1}M(x_1,B_{x_1})}\big).  \label{defNDE}
\end{align}
NIE is also called the mediation effect, and the total causal effect of the independent variable $X$ on outcome $Y$ is equal to the summation of NIE and NDE. 

\begin{figure}
  \begin{center}
  \includegraphics[width =1\textwidth,angle=0]{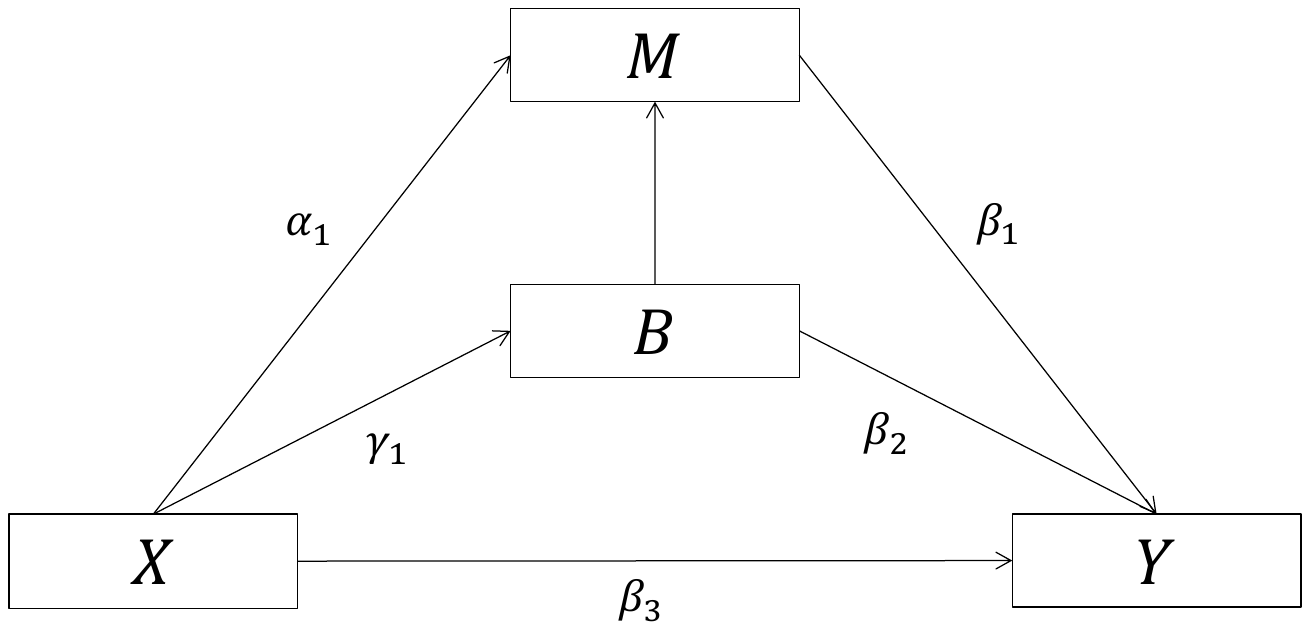}
  \end{center}
  \caption{Potential causal mediation pathways of a zero-inflated mediator.}
  \label{fig_MediatPath}
\end{figure}

The NIE can be further decomposed for a zero-inflated mediator \citep{Steen2017}: 
\begin{align}
\text{NIE}&=E\big(Y_{x_2B_{x_2}M(x_2,B_{x_2})}-Y_{x_2B_{x_1}M(x_1,B_{x_1})}\big) \nonumber\\
&=E\big(Y_{x_2B_{x_2}M(x_2,B_{x_2})} - Y_{x_2B_{x_2}M(x_1,B_{x_1})}\big) + E\big(Y_{x_2B_{x_2}M(x_1,B_{x_1})}  - Y_{x_2B_{x_1}M(x_1,B_{x_1})}\big) \nonumber\\
&\coloneqq \text{NIE}_1+\text{NIE}_2, \label{decomposeNIE}
\end{align}
where NIE$_1$ can be considered as the mediation effect through the numerical change of mediator $M$, and $\text{NIE}_2$ can be considered as the mediation effect through the binary variable $B$ \citep{Steen2017,Daniel2014}. The causal diagram can be seen in Figure \ref{fig_MediatPath}. The NIE, NDE as well as the decomposition of NIE into $\text{NIE}_1$ and $\text{NIE}_2$ in equation (\ref{decomposeNIE}) are identifiable following the assumptions in subsection \ref{assump}.

\section{Model and notation}\label{sc:model}
Notice that the decomposition in equation (\ref{decomposeNIE}) can be applied to general cases such as continuous, binary, count, and survival outcomes, but we focus on continuous outcomes in this paper. For ease of presentation, the subject index is suppressed throughout this section. With a continuous outcome variable $Y$, a zero-inflated mixture mediator $M$, and an independent variable $X$, we construct a full mediation model in this section. First, the following equation will be used to model the association between $Y$ and $(M,B,X)$:
\begin{align}
&Y_{xbm}=\beta_0+\beta_1m+\beta_2 b+\beta_3x+\beta_4xb+\beta_5xm+\epsilon . \label{ymodel}
\end{align}
Equation (\ref{ymodel}) is essentially a regression model for the potential outcome $Y_{xbm}$ allowing for interactions between the independent variable $X$ and the binary mediator $B$ as well as the continuous mediator $M$ respectively, $\epsilon$ is the random error following the normal distribution $N(0,\delta^2)$, and $\beta=(\beta_1,\beta_2,\beta_3,\beta_4,\beta_5)^T$ are the corresponding regression coefficients.

The interactions between $X$ and the two mediators $M$ and $B$ are considered in the model, which is an advantage of potential-outcomes mediation analysis approaches \citep{VanderWeele2009,Imai2010}.
Although for parsimony the model in this section does not include confounders, confounders can be easily incorporated in the model. 

For the zero-inflated mixture mediator $M$, a general form for its two-part density or mass function can be written as:
\begin{align}
&f(m;\mathbf{\theta})=\begin{cases}
    \Delta, & m=0\\
    (1-\Delta)\sum_{k=1}^K\psi_kG_k(m;\theta_k), &m>0
  \end{cases}, \label{zid}
\end{align}
where $0<\Delta<1$ is the probability of $M$ being $0$ (i.e., $P(B=0)=\Delta$), $K$ is the number of mixture components, $\sum_{k=1}^K\psi_kG_k(m;\theta_k)$ is the probability mass or density function of the positive values conditional on $B=1$ (i.e., $M^p | B=1$), $0<\psi_k<1$ is the mixing probability of the $k$th distribution in the mixture such that $\sum_{k=1}^K\psi_k=1$, and $G_k(m;\theta_k)$ is the $k$th component in the mixture with parameter vector $\theta_k$. To model the dependence of $M$ and $B$ on $X$, we can allow the parameters $\Delta$ and $\theta_k$ to depend on $X$. More specifically, we develop zero-inflated log-normal mixture (ZILoNM), zero-inflated Poisson mixture (ZIPM) and zero-inflated negative binomial mixture (ZINBM) mediation models respectively. Again, confounders can be simply added to the model as necessary.

\subsection{Zero-inflated log-normal mixture (ZILoNM) mediators}\label{zilonmSec}
For a mediator $M$ having a zero-inflated log-normal distribution (ZILoNM) with K components, its two-part density function in Equation (\ref{zid}) can be written as:
\begin{align}
&  f(m;\mathbf{\theta})=\begin{cases}
    \Delta, & m=0\\
    (1-\Delta)\sum_{k=1}^K\psi_k\phi_k(m;\mu_k,\sigma), &m>0
  \end{cases}, \label{zilonm}
\end{align}
where the parameter vector $(\mu_k,\sigma)^T$ corresponds to $\theta_k$ in the general equation (\ref{zid}), and $\phi_k(\cdot)$ is the density function of the $k$th log-normal distribution component in the mixture indexed by the parameters $\mu_k$ and $\sigma$ which are the expected value and standard deviation respectively of the random variable after natural-log transformation. For the mixture of $K$ log-normal distributions, the case of equal variances is considered here given that likelihood functions of normal mixtures with unequal variances are not always well-defined \citep{Quandt1978}. Let $M$ and $B$ depend on $X$ through the following equations: 
\begin{align}
&\mu_k=\alpha_{k0}+\alpha_{k1}X, \label{transzilon1}\\
&\log{\bigg(\frac{\Delta}{1-\Delta}\bigg)}=\gamma_0+\gamma_1X.  \label{transzilon2}
\end{align}
Equations (\ref{ymodel}), (\ref{transzilon1}) and (\ref{transzilon2}) together form the full mediation model for a ZILoNM distributed mediator and a continuous outcome. In this paper, we will also construct mediation models for ZIPM and ZINBM distributed mediators respectively (See sections \ref{zipmSec} and \ref{zinbmSec}). 

\subsection{Zero-inflated Poisson mixture (ZIPM) mediators}\label{zipmSec}
Equation (\ref{zid}) for a ZIPM mediator can be rewritten as: 
\begin{align}
&  f(m;\mathbf{\theta})=\begin{cases}
    \Delta, & m=0\\
    (1-\Delta)\sum_{k=1}^K\frac{\psi_k\frac{\exp{(-\lambda_k)}\lambda_k^m}{m!}}{\sum_{k=1}^K\psi_k(1-\exp{(-\lambda_k))}}, & m=1,2,\dots
  \end{cases}, \label{zipm}
\end{align}
where $\Delta=\Delta^*+(1-\Delta^*)\sum_{k=1}^K\psi_k\exp{(-\lambda_k)}$ is the probability of observing a zero, $\lambda_k>0$ is the parameter of the $k$th Poisson distribution in the mixture that controls the amount of zeros from the Poisson mixture distribution, and $0<\Delta^*<1$ is the parameter controlling the amount of excessive $0$'s (i.e., not generated from the Poisson mixture distribution). Let $M$ and $B$ depend on $X$ through the following equations: 
\begin{align}
&\log{(\lambda_k)}=\alpha_{k0}+\alpha_{k1}X , \label{transzip1}\\
&\log{\bigg(\frac{\Delta^*}{1-\Delta^*}\bigg)}=\gamma_0+\gamma_1X. \label{transzip2}
\end{align}
Equations (\ref{ymodel}), (\ref{transzip1}) and (\ref{transzip2}) together form the full mediation model for a ZIPM mediator and a continuous outcome. 

\subsection{Zero-inflated negative binomial mixture (ZINBM) mediators}\label{zinbmSec}
Equation (\ref{zid}) for a ZINBM mediator with K components can be rewritten as: 
\begin{align}
&  f(m;\mathbf{\theta})=\begin{cases}
    \Delta, & m=0\\
    (1-\Delta)\sum_{k=1}^K\frac{\psi_k\frac{\Gamma(r+m)}{\Gamma(r)m!}(\frac{\mu_k}{r+\mu_k})^m(\frac{r}{r+\mu_k})^{r}}{1-\sum_{k=1}^K\psi_k(\frac{r}{r+\mu_k})^{r}}, & m=1,2,\dots
  \end{cases}, \label{zinbm}
\end{align}
where $\Delta=\Delta^*+(1-\Delta^*)\sum_{k=1}^K\psi_k(\frac{r}{r+\mu_k})^{r}$ is the probability of observing a zero, $(\mu_k,r)^T>0$ are the parameters of the $k$th negative binomial distribution in the mixture that controls the amount of zeros generated from the NB mixture distribution, $\mu_k$ is the expectation of the $k$th NB mixture, $r$ is the dispersion parameter that controls amount of over-dispersion, and $0<\Delta^*<1$ is the parameter controlling the amount of excessive zeros in addition to those from the NB mixture. Let $M$ and $B$ depend on $X$ through the following equations:
\begin{align}
&\log{(\mu_k)}=\alpha_{k0}+\alpha_{k1}X , \label{transzinb}\\
&\log{\bigg(\frac{\Delta^*}{1-\Delta^*}\bigg)}=\gamma_0+\gamma_1X. \label{transzinb2}
\end{align}
The dispersion parameter $r$ is assumed to be independent of X and also the same across components in the mixture although it can be allowed to depend on $X$ if needed. Equations (\ref{ymodel}), (\ref{transzinb}) and (\ref{transzinb2}) together form the full mediation model for a ZINBM mediator and a continuous outcome. 

\subsection{Probability mechanism for observing false zeros}\label{pfalse0}
We adapt the mechanism from \cite{Jiang2023}. As in the motivating examples, we often see two types of zeros for $M$ in a data set: true zeros and false zeros. A true zero represents that a measurement is 0, while a false zero means the underlying true value is non-zero but it was observed as a zero due to measurement errors. Denote $M$ as the true value of the mediator and $M^*$ as the observed value of $M$. If the observed value of the mediator is positive $M^*>0$, it is assumed to have been accurately measured (i.e., $M^*=M$). If $M^*=0$, we do not get to observe whether $M$ is truly zero or $M$ is positive but incorrectly observed as a false zero.
\begin{equation*}
  P(M^*=0|M)=\begin{cases}
    \exp(-\eta^2 M), & M\le L\\
    0, &M>L
  \end{cases},
\end{equation*}
where $\eta>0$ is a model parameter, and $L\ge0$ is a known constant. A true zero will be observed as a zero with a probability of 100\%. Under the mechanism, a larger true value is less likely to be observed as a zero and any value above the threshold $L$ cannot be observed as a zero. It is worth  mentioning that there is another commonly used mechanism for false zeros in the literature \citep{Neelon2016} which is called censored zeros (or zeros due to limit of detection) under which the obverved value is always zero if the true value is less than a threshold. Censored zeros do not involve randomness which is the main difference compared with our mechanism.

\subsection{Direct and indirect effects for ZILoNM mediators}\label{effects_zilonm}
Given the proposed model for ZILoNM mediators in subsection \ref{zilonmSec}, we obtain the following formulas by plugging the equations (\ref{ymodel}) and (\ref{zilonm})-(\ref{transzilon2}) into the definitions of effects in section \ref{sc:effects} and then performing Riemann-Stieltjes integration \citep{TerHorst1986}:
\begin{align*}
\text{NIE}_1&=E\big(Y_{x_2B_{x_2}M(x_2,B_{x_2})} - Y_{x_2B_{x_2}M(x_1,B_{x_1})}\big) \nonumber\\
&=(\beta_1+\beta_5x_2)\Bigg\{(1-\Delta_{x_2})\Bigg[\sum_{k=1}^K\psi_k\exp{\big(\mu_{x_2,k}+\frac{\sigma^2}{2}\big)}\Bigg]-(1-\Delta_{x_1})\Bigg[\sum_{k=1}^K\psi_k\exp{\big(\mu_{x_1,k}+\frac{\sigma^2}{2}\big)}\Bigg] \Bigg \},\\
\text{NIE}_2&=E\big(Y_{x_2B_{x_2}M(x_1,B_{x_1})}  - Y_{x_2B_{x_1}M(x_1,B_{x_1})}\big) \nonumber\\
&=(\beta_2+\beta_4x_2)(\Delta_{x_1}-\Delta_{x_2}),\\
\text{NDE}&=E\big(Y_{x_2 B_{x_1}M(x_1,B_{x_1})}-Y_{x_1 B_{x_1}M(x_1,B_{x_1})}\big) \nonumber\\
&=(x_2-x_1)\Bigg\{\beta_3+(1-\Delta_{x_1})\Bigg[\beta_4+\beta_5\bigg(\sum_{k=1}^K\psi_k\exp{\big(\mu_{x_1,k}+\frac{\sigma^2}{2}\big)}\bigg) \Bigg]\Bigg\}.
\end{align*}
More details of the derivation can be found in the Appendix in section \ref{sc:appendix}.

When the distribution of the mediator is not zero-inflated, $\text{NIE}_2$ becomes 0 since $\Delta_x$ reduces to 0, and thus the NIE reduces to a usual NIE that can be calculated by standard approaches \citep{Imai2010, VanderWeele2009}. The derivations of mediation and direct effects for ZIPM and ZINBM can be found in the Appendix in section \ref{sc:appendix}.

\subsection{Assumptions}\label{assump}
For identifiability, the following six assumptions are required for causal interpretation of mediation and direct effects:
\begin{align}
&1.\hspace{0.2cm} Y_{xbm}\ind X | Z,\label{assump1}\\
&2.\hspace{0.2cm}Y_{xbm}\ind \{B_x, M_x\} | X=x,Z,\label{assump2}\\
&3.\hspace{0.2cm}\{B_x, M_x\} \ind X | Z,\label{assump3}\\
&4.\hspace{0.2cm}Y_{xbm}\ind \{B_{x'}, M_{x'}\} | Z,\label{assump4}\\
&5.\hspace{0.2cm}\text{The effect of the first mediator (i.e., $B$) on the second mediator (i.e., $M$)}\nonumber\\
&\hspace{0.6cm}\text{is not confounded within strata of $\{X, Z\}$}\nonumber,\\
&6.\hspace{0.2cm}\text{None of the confounders (if any) for the effect of the first mediator on the }\nonumber\\
&\hspace{0.6cm}\text{second mediator is affected by $X$ }\nonumber,
\end{align}
where $Z$ denotes observed confounders (if any) and $U\ind V | W$ means that random variables $U$ and $V$ are conditionally independent given $W$. These assumptions are typical assumptions and also required by methods such as the MSM method and sequential mediators \citep{VanderWeele2009,multipleMedi14, Steen2017, Daniel2014}. The interpretation of the first three assumptions in equations (\ref{assump1}) - (\ref{assump3}) is that there are no unmeasured confounders for the $X-Y$, $\{B, M\}-Y$ and $X-\{B, M\}$ associations. The fourth assumption in equation (\ref{assump4}) requires that there are no post-baseline confounders of the $\{B, M\}-Y$ association affected by the independent variable $X$.

\section{Estimation}\label{sc:est}
\subsection{Log-likelihood function}
EM algorithm will be used to obtain the MLE of model parameters. We show the derivation for models with ZILoNM mediators. Derivations for ZIPM and ZINBM mediators are provided in the Appendix. As the mediator is a zero-inflated mixture, we define an indicator variable $C$ based on the underlying value of the mediator M as follows:
\begin{align*}
C&=\begin{cases}
    0, &M=0 \hspace{0.1cm}(\text{or}\hspace{0.1cm} B=0)\\
    k, &M\sim\phi_k(\mu_{k},\sigma)
  \end{cases}.
\end{align*}
In other words, $C_i=0$ if the $i$th subject has the true value of $M$ equal to 0, and $C_i=k$ if the $i$th subject belongs to the $k$th log-normal distribution (mixture component) with density $\phi_k(m;\mu_{ik},\sigma)$. Notice that $C_i$ is a latent variable that cannot be observed completely given the mixture model and the possible presence of false zeros. In practice, what we observe instead is $R=1_{(M^*\ne0)}$. Based on the mechanism of observing false zeros in subsection \ref{pfalse0}, we only know individuals with positive mediator values ($R_i=1$) having a value of $C_i\ne0$, but the membership of the log-normal component in the mixture is not observable. For individuals with observed mediator values as zeros ($R_i=0$), it is not known whether they have true zero values ($C_i=0$) or if they were incorrectly observed as false zeros ($C_i\ne0$). Thus, it will be treated as missing data.
The complete data is $(y_i,m_i^*,r_i,c_i,x_i)$, and the complete data log-likelihood function takes the form:
\begin{align*}
\ell&=\log\Big(\prod_{i=1}^Nf(y_i,m_i^*,r_i,c_i|x_i)\Big)\\
&=\log{\Bigg\{\prod_{i=1}^N\prod_{k=0}^K\Big[P(c_i=k)f(y_i,m_i^*,r_i|x_i,c_i=k)\Big]^{1_{(c_i=k)}}\Bigg\}}\\
&=\sum_{i=1}^N\sum_{k=0}^K\Bigg\{1_{(c_i=k)}\Big[\log\big(P(c_i=k)\big)+\log\big(f(y_i,m_i^*,r_i|x_i,c_i=k)\big)\Big]\Bigg\}.
\end{align*}
Let $\Psi_{ik}=P(c_i=k)$ and $\ell_{ik}=\log{\big(f(y_i,m_i^*,r_i|x_i,c_i=k)\big)}$. Then
\begin{align*}
\ell=\sum_{i=1}^N\sum_{k=0}^K1_{(c_i=k)}\Big(\log{(\Psi_{ik})}+\ell_{ik}\Big),
\end{align*}
where
\begin{align*}
\Psi_{ik}&=P(c_i=k)\\
&=\begin{cases}
    P(c_i=0)=P(m_i=0)=\Delta_i, & k=0\\
    P(m_i>0)P(c_i=k|m_i>0)=(1-\Delta_i)\psi_k, &k=1,\dots,K
  \end{cases}.
\end{align*}
The subjects are divided into two groups by whether $M_i^*$, the $i$th individual's observed mediator value, is positive or zero. We use the indicator variable $R_i$ to differentiate these two groups based on observed data. The first group consists of subjects whose observed $M_i$ are positive ($R_i=1$), and the membership in the log-normal mixture is unknown. The second group consists of subjects whose observed mediator value are zeros ($R_i=0$). Furthermore, group 2 has two subgroups: some of them have true zero mediator values ($C_i=0$ and $M_i=0$), while the rest of them were falsely observed as zero values ($C_i\ne0$ and $M_i>0$). The complete data log-likelihood function can be calculated as:
\begin{align}\label{completell}
\ell=\sum_{i\in\text{group1}}\sum_{k=1}^K1_{(c_i=k)}\Big(\log{(\Psi_{ik})}+\ell_{ik}^1\Big)+\sum_{i\in\text{group2}}\sum_{k=0}^K1_{(c_i=k)}\Big(\log{(\Psi_{ik})}+\ell_{ik}^2\Big),
\end{align}
where $\ell_{ik}^1$ and $\ell_{ik}^2$ denote $\ell_{ik}$ in group 1 and 2 respectively.   

The log-likelihood contribution from the $i$th individual from mixture component $k$ in group 1 ($m_i=m_i^*>0$) can be calculated as:
\begin{align*}
\ell_{ik}^1&=\log{\big(f(y_i,m_i^*,r_i=1|x_i,c_i=k)\big)}\\
&=\log(f(y_i, r_i=1|m_i^*,x_i,c_i=k)f(m_i^*|x_i,c_i=k))\\
&=\log(f(y_i|m_i^*,x_i,c_i=k)f(r_i=1|m_i^*,x_i,c_i=k)f(m_i^*|x_i,c_i=k))\\
&=\log(f(y_i|m_i^*,x_i))+\log(f(r_i=1|m_i^*))+\log(f(m_i^*|x_i,c_i=k))\\
&=\log(f(y_i|m_i^*,x_i))+\log(P(M_i^*>0|m_i^*))+\log(f(m_i^*|x_i,c_i=k))\\
&=-\log(\delta)-\frac{(y_i-\beta_0-\beta_1m_i^*-\beta_2-(\beta_3+\beta_4)x_i-\beta_5x_im_i^*)^2}{2\delta^2}+\log\Big[1-1_{(m_i^*\le L)}\exp{(-\eta^2m_i^*)}\Big]\\
&\hspace{0.4cm}-\log(m_i^*\sigma)-\frac{(\log (m_i^*)-\mu_{ik})^2}{2\sigma^2}-\log{(2\pi)}.
\end{align*}

For group 2, the log-likelihood of the individuals with a true zero as their mediator value ($m_i=m_i^*=0$) is
\begin{align*}
\ell_{i0}^2 &= \log(f(y_i,m_i^*=0,r_i=0|x_i,c_i=0))\\
&= \log(f(y_i|m_i^*=0,x_i,c_i=0)f(m_i^*=0,r_i=0|x_i,c_i=0))\\
&= \log(f(y_i|m_i^*=0,x_i)f(r_i=0|m_i^*=0,x_i,c_i=0)f(m_i^*=0|x_i,c_i=0))\\
&= \log(f(y_i|m_i^*=0,x_i)f(r_i=0|m_i^*=0)f(m_i^*=0|x_i,c_i=0))\\
&= \log(f(y_i|m_i^*=0,x_i)\cdot 1 \cdot 1)\\
&=-\log(\delta)-\frac{(y_i-\beta_0-\beta_3x_i)^2}{2\delta^2}-0.5\log{(2\pi)}.
\end{align*}
The rest of group 2 are individuals who truly have non-zero mediator values that were falsely observed as zeros ($m_i>0$ and $m_i^*=0$). Without knowing the true underlying mediator values of such false zeros, integration is used to account for possible non-zero mediator values that are less than or equal to the constant $L$ defined by the mechanism of generating false zeros. The log-likelihood contribution from $k$th mixture component in group 2 can be calculated as:
\begin{align*}
\ell_{ik}^2&=\log(f(y_i,m_i^*=0,r_i=0|x_i,c_i=k))\\
&=\log\bigg(\int\limits_0^L f(y_i, r_i=0|m,x_i,c_i=k)dF(m|x_i,c_i=k)\bigg)\\
&=\log\bigg(\int\limits_0^L f(y_i|m,x_i,c_i=k) f(r_i=0|m,x_i,c_i=k)f(m|x_i,c_i=k)dm\bigg)\\
&=\log\bigg(\int\limits_0^L f(y_i|m,x_i) f(r_i=0|m)\phi(m;\mu_{ik},\sigma)dm\bigg)\\
&=\log\Bigg\{\int\limits_0^L \frac{1}{\delta\sqrt{2\pi}}\exp{\bigg[-\frac{(y_i-\beta_0-\beta_1m-\beta_2-(\beta_3+\beta_4)x_i-\beta_5x_im)^2}{2\delta^2}\bigg]}\\
&\hspace{1.8cm}\cdot \exp{(-\eta^2m)}\frac{1}{m\sigma\sqrt{2\pi}}\exp\bigg[-\frac{(\log (m)-\mu_{ik})^2}{2\sigma^2}\bigg]dm\Bigg\}\\
&=-\log{(\delta)}-\log{(\sigma)}+\log{(h_{ik})}-\log{(2\pi)},
\end{align*}
where
\begin{align*}
h_{ik}=&\int_0^{L}\frac{1}{m}\exp\Bigg(-\frac{(y_i-\beta_0-\beta_1m-\beta_2-(\beta_3+\beta_4)x_i-\beta_5x_im)^2}{2\delta^2}-\eta^2m-\frac{(\log{(m)}-\mu_{ik})^2}{2\sigma^2}\Bigg)dm.\\
\end{align*}
So far, all the terms in $\ell$ in equation (\ref{completell}) are observed except that $1_{(c_i=k)}$ is not observable and constitutes missing data. In order to obtain the MLE of the model parameters, an EM algorithm is constructed to account for the missing data.

\subsection{Expectation step (E step)}
Let $\Theta^0$ denote the initial value of the vector of all of the unknowns. Denote $Q(\Theta|\Theta^0)$ as the expectation of the complete data log-likelihood function with respect to the conditional distribution of the latent variable given data and the current estimates of the model parameters (i.e., distribution of $c_i|y_i,m_i^*,r_i,x_i,\Theta^0$):
\begin{align}\label{logL.Estep}
Q(\Theta|\Theta^0)&=E_{c_i|y_i,m_i^*,r_i,x_i,\Theta^0}(\ell)\nonumber\\
&=\sum_{i\in\text{group1}}\sum_{k=1}^KE_{c_i|y_i,m_i^*,r_i=1,x_i,\Theta^0}\big(1_{(c_i=k)}\big)\Big(\log{(\Psi_{ik})}+\ell_{ik}^1\Big)\nonumber\\
&\hspace{0.4cm}+\sum_{i\in\text{group2}}\sum_{k=0}^KE_{c_i|y_i,m_i^*=0,r_i=0,x_i,\Theta^0}\big(1_{(c_i=k)}\big)\Big(\log{(\Psi_{ik})}+\ell_{ik}^2\Big).
\end{align}
For group 1, let $\tau_{ik}^1(\Theta^0)=E_{c_i|y_i,m_i^*,r_i=1,x_i,\Theta^0}\big(1_{(c_i=k)}\big)$ for $k=1,\dots,K$. By using Bayes formula, we have the following for ZILoNM distributed mediators 
\begin{align*}
\tau_{ik}^1(\Theta^0)&=E\big(1_{(c_i=k)}|y_i,m_i^*,r_i=1,x_i,\Theta^0\big)\\
&=P\big(c_i=k|y_i,m_i^*,r_i=1,x_i,\Theta^0\big)\\
&=\frac{f(y_i,m_i^*,r_i=1,c_i=k|x_i,\Theta^0)}{f(y_i,m_i^*=0,r_i=0|x_i,\Theta^0)}\\
&=\frac{f(y_i,m_i^*,r_i=1|c_i=k,x_i,\Theta^0)P(c_i=k|x_i,\Theta^0)}{\sum_{j=1}^Kf(y_i,m_i^*,r_i=1|c_i=j,x_i,\Theta^0)P(c_i=j|x_i,\Theta^0)}\\
&=\frac{f(y_i|m_i^*,x_i,\Theta^0)f(r_i=1|m_i^*,\Theta^0)f(m_i^*|c_i=k,x_i,\Theta^0)P(c_i=k|x_i,\Theta^0)}{\sum_{j=1}^Kf(y_i|m_i^*,x_i,\Theta^0)f(r_i=1|m_i^*,\Theta^0)f(m_i^*|c_i=j,x_i,\Theta^0)P(c_i=j|x_i,\Theta^0)}\\
&=\frac{f(m_i^*|c_i=k,x_i,\Theta^0)P(c_i=k|x_i,\Theta^0)}{\sum_{j=1}^Kf(m_i^*|c_i=j,x_i,\Theta^0)P(c_i=j|x_i,\Theta^0)}\\
&=\frac{\psi_k^0\phi_k(m_i^*;\mu_{ik}^0,\sigma^0)}{\sum_{j=1}^K\psi_j^0\phi_j(m_i^*;\mu_{ij}^0,\sigma^0)}.
\end{align*}
For group 2, let $\tau_{ik}^2(\Theta^0)=E_{c_i|y_i,m_i^*=0,r_i=0,x_i,\Theta^0}\big(1_{(c_i=k)}\big)$ for $k=0,1,\dots,K$. We have
\begin{align*}
\tau_{ik}^2(\Theta^0)&=E\big(1_{(c_i=k)}|y_i,m_i^*=0,r_i=0,x_i,\Theta^0\big)\\
&=P\big(c_i=k|y_i,m_i^*=0,r_i=0,x_i,\Theta^0\big)\\
&=\frac{f(y_i,m_i^*=0,r_i=0,c_i=k|x_i,\Theta^0)}{f(y_i,m_i^*=0,r_i=0|x_i,\Theta^0)}\\
&=\frac{f(y_i,m_i^*=0,r_i=0|c_i=k,x_i,\Theta^0)P(c_i=k|x_i,\Theta^0)}{\sum_{j=0}^Kf(y_i,m_i^*=0,r_i=0|c_i=j,x_i,\Theta^0)P(c_i=j|x_i,\Theta^0)}\\
&=\frac{\Psi_{ik}\exp{(\ell_{ik}^2)}}{\sum_{j=0}^K\Psi_{ij}\exp{(\ell_{ij}^2)}}\Bigg|_{\hspace{0.1cm}\text{evaluated at}\hspace{0.1cm} \Theta^0}.
\end{align*}
Finally  
\begin{align}\label{logL.Estep2}
Q(\Theta|\Theta^0)&=E(\ell|y_i,m_i^*,r_i,x_i,\Theta^0)\nonumber\\
&=\sum_{i\in\text{group1}}\sum_{k=1}^K\tau_{ik}^1(\Theta^0)\Big(\log{(\Psi_{ik})}+\ell_{ik}^1\Big)+\sum_{i\in\text{group2}}\sum_{k=0}^K\tau_{ik}^2(\Theta^0)\Big(\log{(\Psi_{ik})}+\ell_{ik}^2\Big).
\end{align}

\subsection{Maximization step (M step)}\label{mstep}
We maximize $Q(\Theta|\Theta^0)$ in equation (\ref{logL.Estep2}) with respect to $\Theta$, and find the maximizer $\Theta^1$:  
\begin{align*}
\Theta^1=\underset{\Theta}{\mathrm{arg\;max}}Q(\Theta|\Theta^0)
\end{align*}
The algorithm continues by repeating the E step and M step by replacing $\Theta^0$ with $\Theta^1$ until convergence. The final value of $\Theta^1$ is the MLE.
The standard errors and confidence intervals of the parameter estimates are further calculated using the Fisher information matrix \citep{Oakes1999}. Then, the MLEs of the direct and mediation effects can be easily found using the MLEs of model parameters, and their corresponding standard errors will be determined by the multivariate delta method. Similarly, the log-likelihood functions and an EM algorithm for each of a ZIPM or ZINBM distributed mediator can be found in the Appendix in section \ref{sc:appendix}.

\subsection{Model selection}
In practice, the underlying number of mixture components, $K$, is not known, and thus the estimation procedure needs to select the optimal number of components. We suggest a likelihood-based selection criteria such as the Akaike information criterion (AIC) and the Bayesian information criterion (BIC) since the likelihood function is derived as part of the approach. BIC, compared with AIC, favors more parsimonious models due to heavier penalty on the number of parameters. In the simulation and application sections, we used BIC although AIC generated similar results. We can also let BIC choose the optimal model fit among ZILoNM, ZIPM, and ZINBM while selecting the optimal $K$. These options are available in our R package ``MAZE'', and are implemented in the simulation.

\section{Simulation}\label{sc:simu}
An extensive simulation study was carried out to demonstrate performance of the proposed method comparing to (i) non-mixture method for zero-inflated mediators (i.e., $K=1$) and (ii) a standard causal mediation method, MSM method \citep{VanderWeele2009}. 
In all simulation settings, a total of 100 replications with a sample size of 1000 observations each were generated from a zero-inflated mixture mediator with two mixing components (i.e., $K=2$). 
The proposed method fits a model for $K$ in the range of $1,2,3$ and the three aforementioned distributions (ZILoNM, ZINBM, or ZIPM), and then it selects the final mediation model based on BIC. 
In addition, model parameters were varied to test the model performance under different proportions of zeros including true and false zeros in simulated datasets. Four situations were considered with approximately 30\%, 50\%, 60\%, and 70\% of total zero values in the mediator variable, and these zero values are composed of about half true zeros and half false zeros. 
The independent variable was generated from the standard normal distribution. The average mediation effects including $\text{NIE}_1$, $\text{NIE}_2$, and NIE were calculated for X increasing from $x_1=0$ to $x_2=1$. 
In the simulation, the interaction between the independent variable and mediator $B$ is considered (i.e., $\beta_4\ne0$), and an upper bound of $L=20$ was used in the probability mechanism of observing false zeros based on domain expertise. 

The non-mixture method \citep{Jiang2023} was implemented using the R package ``MAZE'' selecting its final mediation model among ZILoN, ZINB, and ZIP using BIC. To implement the MSM method, the R package ``CMAverse'' \citep{Shi2021} was applied with two mediators $M$ and $B$. Since MSM is not designed to handle excessive zero mediator values, only an average NIE as opposed to $\text{NIE}_1$ and $\text{NIE}_2$ could be estimated. 

\subsection{Data generated from ZILoNM mediators}
When data were simulated from 2-component ZILoNM mediators, the results in Table \ref{table_simzilonm} showed that the proposed method had negligible percentage bias and coverage probability close to the nominal level of 0.95, which indicates good estimations of average mediation effects and its standard deviation. In addition, it is robust to the presence of various percentages of zero values. The true distribution of the mediator (ZILoNM) and number of mixing components ($K=2$) were chosen for over 99\% of the generated datasets. When a mis-specified model was selected occasionally, its results were still included as an assessment to model robustness. 

For the non-mixture method (i.e., $K=1$), the estimated $\text{NIE}_1$ was not accurate enough since a single log-normal distribution cannot properly assemble the mixture distribution, and it became worse with increasing number of zero mediator values. A similar trend was observed for performance of the estimators of $\text{NIE}_2$ and total NIE, but they tend to have smaller biases at the presence of fewer zeros. 

Since the existing software for the MSM method does not offer the option for ZILoN mediators (including the R package ``CMAverse'' \citep{Shi2021}), the comparison is not included in this setting.

\begin{table}
\caption{Simulation results$^+$ for 2-component ZILoNM data when $X$ changes from $x_1=0$ to $x_2=1$ for the proposed method versus the non-mixture method. Mean SE is the mean of estimated standard errors. Bias is the mean estimate minus the corresponding true value, and percent of bias is the bias as a percentage of true values. CP is the empirical coverage probability of the 95\% CI.} 
\label{table_simzilonm}
\centering
\resizebox{14cm}{!}{
\begin{tabular}{c c c c c c c c c} \\
\hline
Total & Method & Mediation & True & Mean & Mean & Bias & Percent of Bias & CP\\
zeros (\%) & & Effect && Estimate & SE && (\%) & \\
\hline\hline\\
\multirow{1}{3em}{$\sim 30$} & Proposed & $\text{NIE}_1$ & 0.07 & 0.07 & 0.02 & 0.00 & 1.52 & 0.95 \\ 
& 100\% selected ZILoNM $K=2$ & $\text{NIE}_2$ & 0.75 & 0.75 & 0.17 & -0.01 & -0.66 & 0.96 \\ 
&& NIE & 0.82 & 0.81 & 0.16 & -0.00 & -0.49 & 0.96 \\ \\
& Non-mixture (i.e., $K=1$) & $\text{NIE}_1$ & 0.07 & 0.05 & 0.01 & -0.02 & -25.76 & 0.78 \\ 
&   & $\text{NIE}_2$ & 0.75 & 0.76 & 0.17 & 0.01 & 1.06 & 0.96 \\ 
&& NIE & 0.82 & 0.81 & 0.16 & -0.01 & -1.10 & 0.96 \\ 
\hline \\
\multirow{1}{3em}{$\sim 50$} & Proposed & $\text{NIE}_1$ & 0.04 & 0.04 & 0.01 & 0.00 & 4.76 & 0.96 \\ 
& 100\% selected ZILoNM $K=2$ & $\text{NIE}_2$ & 1.09 & 1.09 & 0.20 & 0.00 & 0.00 & 0.97 \\ 
&& NIE & 1.13 & 1.13 & 0.20 & 0.00 & 0.18 & 0.97 \\  \\
& Non-mixture (i.e., $K=1$) & $\text{NIE}_1$ & 0.04 & -0.10 & 0.03 & -0.14 & -345.24 & 0.58 \\ 
&  & $\text{NIE}_2$ & 1.09 & 0.76 & 0.16 & -0.33 & -29.96 & 0.71 \\ 
&& NIE & 1.13 & 0.66 & 0.17 & -0.47 & -41.68 & 0.71 \\ 
\hline \\
\multirow{1}{3em}{$\sim 60$} & Proposed & $\text{NIE}_1$ & 0.03 & 0.04 & 0.01 & 0.00 & 0.00 & 0.97 \\ 
& 100\% selected ZILoNM $K=2$ & $\text{NIE}_2$ & 1.17 & 1.22 & 0.21 & 0.04 & 3.66 & 0.95 \\ 
&& NIE & 1.21 & 1.25 & 0.20 & 0.04 & 3.64 & 0.95 \\  \\
& Non-mixture (i.e., $K=1$) & $\text{NIE}_1$ & 0.03 & -0.39 & 0.07 & -0.42 & -1211.43 & 0.17 \\ 
&  & $\text{NIE}_2$ & 1.17 & 0.11 & 0.11 & -1.07 & -90.89 & 0.16 \\ 
&& NIE & 1.21 & -0.28 & 0.12 & -1.49 & -123.33 & 0.16 \\ 
\hline \\
\multirow{1}{3em}{$\sim 70$} & Proposed & $\text{NIE}_1$ & 0.03 & 0.03 & 0.01 & 0.00 & 0.00 & 0.91 \\ 
& 99\% selected ZILoNM $K=2$ & $\text{NIE}_2$  & 1.24 & 1.28 & 0.22 & 0.04 & 3.30 & 0.92 \\ 
&& NIE & 1.28 & 1.32 & 0.21 & 0.04 & 3.29 & 0.92 \\   \\
& Non-mixture (i.e., $K=1$) & $\text{NIE}_1$ & 0.03 & -0.45 & 0.12 & -0.49 & -1435.29 & 0.05 \\ 
&  & $\text{NIE}_2$ & 1.24 & 0.55 & 0.21 & -0.69 & -55.55 & 0.16 \\ 
&& NIE & 1.28 & 0.10 & 0.22 & -1.18 & -92.25 & 0.07 \\
\hline
\multicolumn{9}{l}{Proposed: proposed method}\\
\multicolumn{9}{l}{$\text{NIE}_1$: mediation effect attributable to the change in the mediator on the numerical scale}\\
\multicolumn{9}{l}{$\text{NIE}_2$: mediation effect attributable to the binary change of the mediator from zero to a non-zero status}\\
\multicolumn{9}{l}{NIE: total mediation effect}\\
\multicolumn{9}{l}{$^+$Results for the MSM method were not included as the existing R package is not applicable}\\[1ex]
\end{tabular}
}
\end{table}

\subsection{Data generated from ZINBM mediators}
As shown in Table \ref{table_simzinbm}, good performance of the proposed approach was observed in terms of both biases and coverage probability at different proportions of zeros. As the number of zeros increases, less information was available for the proposed method to proceed with estimation, so fewer final models with the true distribution of mediator and number of components $K$ were chosen. Nevertheless, when the true model was not selected, the final performance stays reasonable.

The non-mixture method (i.e., $K=1$) had acceptable results for $\text{NIE}_1$ when there was a small number of zero mediator values  possibly due to the fact that a NB distribution itself is already a mixture of Poisson distributions, which has good flexibility in terms of over-dispersion distribution shape. With more zeros, all estimates had larger biases and decreased CP.

In contrast, the MSM method had undesirable performance in all settings, which is possibly caused by the failure to account for the mixture structure of non-zero mediator values in addition to the presence of excessive zeros. 

\begin{table}
\caption{Simulation results for 2-component ZINBM data when $X$ changes from $x_1=0$ to $x_2=1$ for the proposed method versus the non-mixture method and MSM. Mean SE is the mean of estimated standard errors. Bias is the mean estimate minus the true parameter value, and percent of bias is the bias as a percentage of true values. CP is the empirical coverage probability of the 95\% CI.
There are about 1.5\% zeros from the NB mixture among the true zeros in each case. }
\label{table_simzinbm}
\centering
\resizebox{14cm}{!}{
\begin{tabular}{c c c c c c c c c} \\
\hline
Total & Method & Mediation & True & Mean & Mean & Bias & Percent of Bias & CP\\
zeros (\%) & & Effect && Estimate & SE && (\%) & \\
\hline\hline\\
\multirow{1}{3em}{$\sim 30$} & Proposed & $\text{NIE}_1$ & -0.52 & -0.54 & 0.15 & -0.03 & 5.42 & 0.94 \\ 
& 99\% selected ZINBM $K=2$ & $\text{NIE}_2$ & -0.45 & -0.43 & 0.19 & 0.02 & -4.26 & 0.96 \\ 
&& NIE & -0.96 & -0.97 & 0.22 & -0.01 & 0.83 & 0.93 \\  \\
& Non-mixture (i.e., $K=1$) & $\text{NIE}_1$ & -0.52 & -0.56 & 0.16 & -0.05 & 9.09 & 0.94 \\  
&  & $\text{NIE}_2$ & -0.45 & -0.34 & 0.16 & 0.11 & -24.22 & 0.61 \\ 
&& NIE & -0.96 & -0.90 & 0.22 & 0.06 & -6.33 & 0.71 \\  \\
& MSM & NIE & -0.96 & -0.55 & 0.19 & 0.42 & -43.30 & 0.36 \\ 
\hline \\
\multirow{1}{3em}{$\sim 50$} & Proposed & $\text{NIE}_1$  & -0.36 & -0.37 & 0.13 & -0.00 & 1.10 & 0.94 \\ 
& 92\% selected ZINBM $K=2$  & $\text{NIE}_2$  & -0.69 & -0.69 & 0.30 & 0.00 & -0.43 & 0.90 \\
&& NIE & -1.06 & -1.06 & 0.30 & -0.00 & 0.09 & 0.92 \\  \\
& Non-mixture (i.e., $K=1$) & $\text{NIE}_1$  & -0.36 & -0.34 & 0.13 & 0.02 & -5.79 & 0.92 \\
&  & $\text{NIE}_2$ & -0.69 & -0.94 & 0.28 & -0.24 & 35.16 & 0.89 \\ 
&& NIE & -1.06 & -1.28 & 0.31 & -0.22 & 21.00 & 0.87 \\   \\
& MSM & NIE & -1.06 & -0.41 & 0.19 & 0.65 & -61.31 & 0.01 \\ 
\hline \\
\multirow{1}{3em}{$\sim 60$} & Proposed & $\text{NIE}_1$ & -0.30 & -0.29 & 0.13 & 0.01 & -3.70 & 0.95 \\ 
& 93\% selected ZINBM $K=2$ & $\text{NIE}_2$ & -0.78 & -0.76 & 0.35 & 0.02 & -2.56 & 0.90 \\ 
& & NIE & -1.08 & -1.05 & 0.34 & 0.03 & -2.79 & 0.93 \\    \\
& Non-mixture (i.e., $K=1$) & $\text{NIE}_1$ & -0.30 & -0.26 & 0.12 & 0.04 & -12.79 & 0.92 \\ 
&  & $\text{NIE}_2$ & -0.78 & -1.06 & 0.32 & -0.28 & 36.03 & 0.89 \\ 
&& NIE & -1.08 & -1.32 & 0.34 & -0.24 & 22.56 & 0.88 \\   \\
& MSM & NIE & -1.08 & -0.26 & 0.18 & 0.81 & -75.58 & 0.00 \\ 
\hline \\
\multirow{1}{3em}{$\sim 70$} & Proposed & $\text{NIE}_1$ & -0.23 & -0.24 & 0.13 & -0.00 & 1.70 & 0.90 \\ 
& 78\% selected ZINBM $K=2$ & $\text{NIE}_2$ & -0.85 & -0.82 & 0.37 & 0.03 & -3.31 & 0.91 \\ 
&& NIE & -1.08 & -1.06 & 0.36 & 0.02 & -2.22 & 0.91 \\ \\
& Non-mixture (i.e., $K=1$) & $\text{NIE}_1$ & -0.23 & -0.21 & 0.12 & 0.02 & -8.51 & 0.87 \\ 
&  & $\text{NIE}_2$ & -0.85 & -1.16 & 0.39 & -0.31 & 37.04 & 0.92 \\ 
&& NIE & -1.08 & -1.37 & 0.40 & -0.29 & 27.13 & 0.89 \\  \\
& MSM & NIE & -1.08 & -0.26 & 0.17 & 0.82 & -75.65 & 0.00 \\ 
\hline
\multicolumn{9}{l}{Proposed: proposed method}\\
\multicolumn{9}{l}{MSM: marginal structural models}\\
\multicolumn{9}{l}{$\text{NIE}_1$: mediation effect attributable to the change in the mediator on the numerical scale}\\
\multicolumn{9}{l}{$\text{NIE}_2$: mediation effect attributable to the binary change of the mediator from zero to a non-zero status}\\
\multicolumn{9}{l}{NIE: total mediation effect}\\[1ex]
\end{tabular}
}
\end{table}

\subsection{Data generated from ZIPM mediators}
As shown in Table \ref{table_simzipm}, the mediation effects from the proposed method were estimated with small biases, and the 95\% confidence intervals have reasonable coverage probability across different proportions of zero mediator values. The true model with 2-component ZIPM mediators was chosen for more than 99\% of generated datasets.

The non-mixture method for ZIPM data produced reasonably good estimates and standard errors at different proportions of zero values possibly due to the fact that a NB distribution is a mixture of Poisson distributions.

On the other hand, the bias and coverage probability of the estimation in the MSM method are far from ideal especially when the number of zeros increases. 

\begin{table}
\caption{Simulation results for 2-component ZIPM data when $X$ changes from $x_1=0$ to $x_2=1$ for the proposed method versus the non-mixture method and MSM. Mean SE is the mean of estimated standard errors. Bias is the mean estimate minus the true parameter value, and percent of bias is the bias as a percentage of true values. CP is the empirical coverage probability of the 95\% CI.
There are about 5\% zeros from the Poisson mixture among the true zeros in each case.} 
\label{table_simzipm}
\centering
\resizebox{14cm}{!}{
\begin{tabular}{c c c c c c c c c} \\
\hline
Total & Method & Mediation & True & Mean & Mean & Bias & Percent of Bias & CP\\
zeros (\%) & & Effect && Estimate & SE && (\%) & \\
\hline\hline\\
\multirow{1}{3em}{$\sim 30$} & Proposed & $\text{NIE}_1$ & -0.88 & -0.90 & 0.28 & -0.02 & 2.04 & 0.97 \\ 
& 100\% selected ZIPM $K=2$ & $\text{NIE}_2$ & -0.28 & -0.26 & 0.15 & 0.02 & -5.71 & 0.95 \\ 
&& NIE & -1.16 & -1.16 & 0.32 & -0.00 & 0.26 & 0.94 \\ \\
& Non-mixture (i.e., $K=1$) & $\text{NIE}_1$ & -0.88 & -0.90 & 0.28 & -0.02 & 2.49 & 0.95 \\ 
&  & $\text{NIE}_2$ & -0.28 & -0.27 & 0.15 & 0.01 & -2.50 & 0.93 \\ 
&& NIE & -1.16 & -1.18 & 0.33 & -0.02 & 1.38 & 0.94 \\  \\
& MSM & NIE & -1.16 & -0.32 & 0.25 & 0.84 & -72.72 & 0.06 \\ 
\hline \\
\multirow{1}{3em}{$\sim 50$} & Proposed & $\text{NIE}_1$ & -0.69 & -0.70 & 0.26 & -0.01 & 1.44 & 0.94 \\ 
& 100\% selected ZIPM $K=2$ & $\text{NIE}_2$ & -0.60 & -0.59 & 0.26 & 0.01 & -1.34 & 0.94 \\ 
&& NIE & -1.29 & -1.29 & 0.39 & -0.00 & 0.15 & 0.93 \\  \\
& Non-mixture (i.e., $K=1$) & $\text{NIE}_1$ & -0.69 & -0.71 & 0.26 & -0.02 & 2.74 & 0.94 \\ 
&  & $\text{NIE}_2$ & -0.60 & -0.62 & 0.26 & -0.03 & 4.36 & 0.94 \\ 
&& NIE & -1.29 & -1.34 & 0.39 & -0.04 & 3.41 & 0.93 \\  \\
& MSM & NIE & -1.29 & 0.10 & 0.23 & 1.39 & -107.36 & 0.00 \\
\hline \\
\multirow{1}{3em}{$\sim 60$} & Proposed & $\text{NIE}_1$ & -0.64 & -0.61 & 0.25 & 0.03 & -4.06 & 0.98 \\ 
& 99\% selected ZIPM $K=2$ & $\text{NIE}_2$ & -0.67 & -0.63 & 0.29 & 0.04 & -5.98 & 0.93 \\
&& NIE & -1.31 & -1.24 & 0.42 & 0.07 & -5.04 & 0.90 \\   \\
& Non-mixture (i.e., $K=1$) & $\text{NIE}_1$ & -0.64 & -0.63 & 0.25 & 0.01 & -2.18 & 0.98 \\ 
&  & $\text{NIE}_2$ & -0.67 & -0.69 & 0.29 & -0.02 & 2.69 & 0.94 \\ 
&& NIE  & -1.31 & -1.31 & 0.43 & -0.00 & 0.23 & 0.93 \\  \\
& MSM & NIE & -1.31 & 0.12 & 0.21 & 1.43 & -109.39 & 0.00 \\ 
\hline \\
\multirow{1}{3em}{$\sim 70$} & Proposed & $\text{NIE}_1$ & -0.40 & -0.37 & 0.23 & 0.03 & -7.56 & 0.93 \\ 
& 100\% selected ZIPM $K=2$ & $\text{NIE}_2$ & -0.70 & -0.64 & 0.33 & 0.07 & -9.42 & 0.95 \\ 
&  & NIE & -1.10 & -1.00 & 0.44 & 0.10 & -8.74 & 0.90 \\  \\
& Non-mixture (i.e., $K=1$) & $\text{NIE}_1$ & -0.40 & -0.37 & 0.23 & 0.03 & -6.30 & 0.94 \\ 
&  & $\text{NIE}_2$ & -0.70 & -0.73 & 0.31 & -0.03 & 4.28 & 0.92 \\ 
&& NIE & -1.10 & -1.10 & 0.43 & -0.00 & 0.46 & 0.92 \\   \\
& MSM & NIE & -1.10 & -0.06 & 0.18 & 1.04 & -94.44 & 0.00 \\ 
\hline
\multicolumn{9}{l}{Proposed: proposed method}\\
\multicolumn{9}{l}{MSM: marginal structural models}\\
\multicolumn{9}{l}{$\text{NIE}_1$: mediation effect attributable to the change in the mediator on the numerical scale}\\
\multicolumn{9}{l}{$\text{NIE}_2$: mediation effect attributable to the binary change of the mediator from zero to a non-zero status}\\
\multicolumn{9}{l}{NIE: total mediation effect}\\[1ex]
\end{tabular}
}
\end{table}

\section{Real data applications - ABCD Study}\label{sc:application}
To explore the effect of children's behavior problems on crystallized cognition mediated by brain structural connectivities, the proposed method was applied to the ABCD data in comparison with the non-mixture method. 
There are 52\% boys and 48\% girls in the dataset with 8749 children in total. These children were around 10 years old on average at the time of interviewing or sampling, and 84\% of their parents have at least college education level.
There are 7 areas of interests for the independent variable: children's behavior problems assessed as CBCL scores in depression (depress), anxiety disorder (anxdisord), somatic problems (somaticpr), attention-deficit/hyperactivity disorder (adhd), oppositional defiant disorder (opposit), conduct problems (conduct), and obsessive-compulsive problems (ocd).
The outcome variable was crystallized cognition that represents the knowledge and skills acquired through learning and experience.
The two potential mediators, (1) connectivity between right language network and left accumbens area, denoted by LRAL and (2) connectivity between left auditory network and right cerebellum, denoted by ALCR, have 29\% and 34\% of zeros respectively.

We considered one mediator and one independent variable at a time. In analyses, the age, sex, and parent's highest level of education were adjusted as confounders, and the mediation effect was estimated conditional on their average values. 

Since the mediator is a continuous variable in terms of brain structural connectivity, the proposed method only employed ZILoNM mediators as ZIPM and ZINBM are for count data. BIC was used to determine $K$. The bound $L=20$ in the probability mechanism for observing false zeros was used, and the interaction between the independent variable and the binary indicator of mediator greater than zero was included in the outcome model. Similarly, the non-mixture method for ZILoN mediators was used given the continuous mediator variable. The MSM method is not included in the comparison because existing software cannot accommodate ZILoN mediators for MSM.

The selected results before adjustment for multiple testing are shown in Table \ref{table_real_abcd}. The brain structural connectivities between right language network and left accumbens was found to significantly mediate the effect of CBCL scores in oppositional defiant disorder and obsessive-compulsive problems on cognition through NIE$_2$ and total NIE in a selected 2-component mixture model, while the non-mixture method failed to detect it. For the mediation path from CBCL scores of conduct problems on 
cognition through brain structural connectivities between left auditory network and right cerebellum, both methods demonstrate a significant relationship in total NIE, and the proposed method revealed an additional significant NIE$_1$. 

The interpretation of NIE, NIE$_1$, and NIE$_2$ is straightforward. For example, the analysis result showed that the NIE for the potential causal pathway $\text{conduct}\rightarrow \text{ALCR}\rightarrow \text{cognition}$ is -0.05 ($p=0.02$). That means when the conduct score increases from its minimum to the third quantile, the crystallized cognition score decreases by 0.05 through the resulting change of mediator ALCR. The estimated values NIE$_1=-0.04$  ($p=0.04)$ and NIE$_2=-0.01$  ($p=0.34)$ show that NIE can be decomposed into NIE$_1$ and NIE$_2$ where NIE$_1$ is the mediation effect through the numerical change on the continuum scale of the mediator ALCR and NIE$_2$ is the mediation effect through the binary change (i.e., zero to non-zero status) of the mediator ALCR.

\begin{table}
\caption{ABCD Study: results$^+$ for the estimation of mediation effects when children behavior problems change from its minimum to the third quartile for the proposed versus non-mixture method.} 
\label{table_real_abcd}
\centering
\resizebox{14cm}{!}{
\begin{tabular}{c c c c c c c c} \\
\hline
X & M & Method & Mediation & Estimate & SE & 95\% CI & p-value\\
(CBCL scores) &(Brain connectivity)&& Effect &&&&\\
\hline\hline
opposit & LRAL & Proposed & NIE1 &  0.00 & 0.01 & (-0.02,  0.01) & 0.60 \\ 
   &  & ZILoNM K = 2 & NIE2 & -0.07 & 0.03 & (-0.14,  0.00) & 0.04 \\ 
   &  &  & NIE & -0.07 & 0.04 & (-0.14,  0.00) & 0.04 \\ 
   &  & Non-mixture (i.e., K = 1) & NIE1 &  0.00 & 0.01 & (-0.01,  0.02) & 0.84 \\ 
   &  & ZILoN & NIE2 & -0.01 & 0.13 & (-0.27,  0.24) & 0.92 \\ 
   &  &  & NIE & -0.01 & 0.14 & (-0.28,  0.25) & 0.93 \\ 
   \hline
conduct & ALCR & Proposed & NIE1 & -0.04 & 0.02 & (-0.06,  0.00) & 0.04 \\ 
   &  & ZILoNM K = 2 & NIE2 & -0.01 & 0.02 & (-0.05,  0.02) & 0.34 \\ 
   &  &  & NIE & -0.05 & 0.02 & (-0.09, -0.01) & 0.02 \\ 
   &  & Non-mixture (i.e., K = 1) & NIE1 & -0.03 & 0.02 & (-0.06,  0.00) & 0.07 \\ 
   &  & ZILoN & NIE2 & -0.01 & 0.01 & (-0.03,  0.01) & 0.40 \\ 
   &  &  & NIE & -0.04 & 0.02 & (-0.07, -0.01) & 0.02 \\ 
   \hline
ocd & LRAL & Proposed & NIE1 &  0.00 & 0.00 & (-0.01, 0.00) & 0.22 \\ 
   &  & ZILoNM K = 2 & NIE2 &  0.04 & 0.02 & ( 0.01, 0.07) & 0.01 \\ 
   &  &  & NIE &  0.04 & 0.01 & ( 0.02, 0.06) & 0.00 \\ 
   &  & Non-mixture (i.e., K = 1) & NIE1 &  0.00 & 0.01 & (-0.02, 0.01) & 0.55 \\ 
   &  & ZILoN & NIE2 &  0.04 & 0.09 & (-0.15, 0.22) & 0.69 \\ 
   &  &  & NIE &  0.03 & 0.09 & (-0.15, 0.22) & 0.73 \\ 
   \hline
\multicolumn{8}{l}{Proposed: proposed method for ZILoNM mediators chosen by BIC}\\
\multicolumn{8}{l}{$\text{NIE}_1$: mediation effect attributable to the change in the mediator on the numerical scale}\\
\multicolumn{8}{l}{$\text{NIE}_2$: mediation effect attributable to the binary change of the mediator from zero to a non-zero status}\\
\multicolumn{8}{l}{NIE: total mediation effect}\\
\multicolumn{8}{l}{$^+$Results for the MSM method were not included as the existing R package is not applicable}\\[1ex]
\end{tabular}
}
\end{table}

\section{Discussion}\label{sc:diss}
This paper has proposed a novel mediation approach for zero-inflated mixture mediator data where confounders can be easily adjusted. ZILoNM, ZIPM, and ZINBM are considered in the approach to model such mediators. This new approach allows extended flexibility to model the atypical shape of zero-inflated mediator data through mixture distributions, instead of being confined to common distributions. Observing false zeros is accounted for through a probability mechanism. The estimated mediation effect can be decomposed into two parts. One part of the mediation effect comes from the change in mediator on the continuous scale, and the other part results from the change of mediator from zero to a non-zero status. 

In model estimation, a challenge arises from unknown component membership in the mixture for non-zero values; it is not known whether an observed zero is a true or false zero. We overcome this challenge by using the EM algorithm to obtain unbiased estimates and valid inference of the mediation effects. Both the simulation and real data application demonstrated satisfactory performance of the proposed model. The optimal number of components in the mixture is determined by the model selection criterion BIC. The results also demonstrate that the proposed model outperforms the standard method of mediation analysis that ignores zero-inflation. Our proposed method has been incorporated into the R package ``MAZE'' at \url{https://github.com/meilinjiang/MAZE} with options to adjust for confounders.

There are several possible extensions can be made to the proposed approach. The linear model for $Y$ can be extended to a generalized linear model to account for more outcome data types. The probability mechanism for observing false zeros plays an important role in the approach. A possible extension is to incorporate different mechanisms for observing false zeros. It is also worth mentioning the possible extension to multiple or high-dimensional zero-inflated mediators given that many instances of zero-inflation in single-cell sequencing data and microbiome data.

\section*{Acknowledgments}
This research work was partly supported by NIH grants: P01AA029543, R01AG068128 and R01MH124106.

\subsection*{Data availability}
Research data are not shared.

\subsection*{Conflict of interest}
The authors declare no potential conflict of interests.

\appendix
\section{} \label{sc:appendix}
\subsection{Detailed derivations for ZILoNM mediators}
\subsubsection{Mediation effect and direct effect}
\begin{align*}
\text{NIE}_1&=E\big(Y_{x_2B_{x_2}M(x_2,B_{x_2})} - Y_{x_2B_{x_2}M(x_1,B_{x_1})}\big) \\
&=E\Big[E\Big(Y_{x_2 B_{x_2}M(x_2,B_{x_2})}|M(x_2,B_{x_2}),B_{x_2}\Big)\Big]-E\Big[E\Big(Y_{x_2 B_{x_2}M(x_1,B_{x_1})}|M(x_1,B_{x_1}), B_{x_2}\Big)\Big] \\ 
&=E\Big(\beta_0+\beta_1M(x_2,B_{x_2})+\beta_2 B_{x_2}+\beta_3x_2+\beta_4x_2 M_{x_2}+\beta_5x_2M(x_2,B_{x_2})\Big) \\
&\hspace{0.5cm}-E\Big(\beta_0+\beta_1M(x_1,B_{x_1})+\beta_2 B_{x_2}+\beta_3x_2+\beta_4x_2 B_{x_2}+\beta_5x_2M(x_1,B_{x_1})\Big) \\
&=(\beta_1+\beta_5x_2)\Big(E(M(x_2,B_{x_2}))-E(M(x_1,B_{x_1}))\Big)\\
&=(\beta_1+\beta_5x_2)\Bigg[\int\limits_{m\in\Omega}mdF_{M(x_2,B_{x_2})}(m)-\int\limits_{m\in\Omega}mdF_{M(x_1,B_{x_1})}(m)\Bigg]\\
&=(\beta_1+\beta_5x_2)\Bigg[(1-\Delta_{x_2})\int\limits_{m\in\Omega\setminus 0}mf(m;x_2),\sigma)dm-(1-\Delta_{x_1})\int\limits_{m\in\Omega\setminus 0}mf(m;x_1),\sigma)dm\Bigg],\\
&=(\beta_1+\beta_5x_2)\Bigg\{(1-\Delta_{x_2})\Bigg[\sum_{k=1}^K\psi_k\exp{\big(\mu_{x_2,k}+\frac{\sigma^2}{2}\big)}\Bigg]-(1-\Delta_{x_1})\Bigg[\sum_{k=1}^K\psi_k\exp{\big(\mu_{x_1,k}+\frac{\sigma^2}{2}\big)}\Bigg] \Bigg \},\\
\text{NIE}_2&=E\big(Y_{x_2B_{x_2}M(x_1,B_{x_1})}  - Y_{x_2B_{x_1}M(x_1,B_{x_1})}\big)\\
&=E\Big[E\Big(Y_{x_2 B_{x_2}M(x_1,B_{x_1})}|M(x_1,B_{x_1}),B_{x_2}\Big)\Big]-E\Big[E\Big(Y_{x_2 B_{x_1}M(x_1,B_{x_1})}|M(x_1,B_{x_1}),B_{x_1}\Big)\Big] \\ 
&=E\Big(\beta_0+\beta_1M(x_1,B_{x_1})+\beta_2 B_{x_2}+\beta_3x_2+\beta_4x_2B_{x_2}+\beta_5x_2M(x_1,B_{x_1})\Big) \\
&\hspace{0.5cm}-E\Big(\beta_0+\beta_1M(x_1,B_{x_1})+\beta_2 B_{x_1}+\beta_3x_2+\beta_4x_2B_{x_1}+\beta_5x_2M(x_1,B_{x_1})\Big) \\
&=(\beta_2+\beta_4x_2)(E(B_{x_2})-E(B_{x_1}))\\
&=(\beta_2+\beta_4x_2)(\Delta_{x_1}-\Delta_{x_2}),\\
\text{NDE}&=E\big(Y_{x_2 B_{x_1}M(x_1,B_{x_1})}-Y_{x_1 B_{x_1}M(x_1,B_{x_1})}\big) \\
&=\beta_3(x_2-x_1)+\beta_4(x_2-x_1)(1-\Delta_{x_1})+\beta_5(x_2-x_1)E(M(x_1,B_{x_1}))\\
&=(x_2-x_1)\Bigg\{ \beta_3+\beta_4(1-\Delta_{x_1})+\beta_5(1-\Delta_{x_1})\Bigg[\sum_{k=1}^K\psi_k\exp{\big(\mu_{x_1,k}+\frac{\sigma^2}{2}\big)}\Bigg]\Bigg\}\\
&=(x_2-x_1)\Bigg\{\beta_3+(1-\Delta_{x_1})\Bigg[\beta_4+\beta_5\bigg(\sum_{k=1}^K\psi_k\exp{\big(\mu_{x_1,k}+\frac{\sigma^2}{2}\big)}\bigg) \Bigg]\Bigg\},
\end{align*}
where $\Omega$ denotes the domain of the mediator $M$, $\Omega\setminus 0$ denotes the subset of $\Omega$ that does not contain $0$, $F_{M(x,B_{x})}(m)$ denotes the cumulative distribution function of $M(x,B_{x})$, and $dF_{M(x,B_{x})}(m)$ denotes the Stieltjes integration \citep{TerHorst1986} with respect to $F_{M(x,B_{x})}(m)$.

\subsection{Model for zero-inflated Poisson mixture (ZIPM) mediators}
\subsubsection{Mediation effect and direct effect}
In the case of ZIPM mediators, we obtain the following formulas by plugging the equations (\ref{ymodel}) and (\ref{zipm})-(\ref{transzip2}) into definitions of average NIE and NDE for $X$ changing from $x_1$ to $x_2$ in equations (\ref{defNIE})-(\ref{defNDE}), 
\begin{align*}
\text{NIE}&=E\big(Y_{x_2B_{x_2}M(x_2,B_{x_2})}-Y_{x_2B_{x_1}M(x_1,B_{x_1})}\big) \\
&=E\big(Y_{x_2B_{x_2}M(x_2,B_{x_2})} - Y_{x_2B_{x_2}M(x_1,B_{x_1})}\big) + E\big(Y_{x_2B_{x_2}M(x_1,B_{x_1})}  - Y_{x_2B_{x_1}M(x_1,B_{x_1})}\big) \\
&=\text{NIE}_1+\text{NIE}_2,\\
\text{NIE}_1&=E\big(Y_{x_2B_{x_2}M(x_2,B_{x_2})} - Y_{x_2B_{x_2}M(x_1,B_{x_1})}\big) \\
&=(\beta_1+\beta_5x_2)\Big(E(M(x_2,B_{x_2}))-E(M(x_1,B_{x_1}))\Big)\\
&=(\beta_1+\beta_5x_2)\Bigg[\sum_{m=0}^\infty mf(m;x_2)-\sum_{m=0}^\infty mf(m;x_1)\Bigg]\\
&=(\beta_1+\beta_5x_2) \Bigg[(1-\Delta^*_{x_2})\sum_{k=1}^K\psi_k\lambda_{k,x_2}-(1-\Delta^*_{x_1})\sum_{k=1}^K\psi_k\lambda_{k,x_1}\Bigg],\\
\text{NIE}_2&=E\big(Y_{x_2B_{x_2}M(x_1,B_{x_1})}  - Y_{x_2B_{x_1}M(x_1,B_{x_1})}\big)\\
&=(\beta_2+\beta_4x_2)(E(B_{x_2})-E(B_{x_1}))\\
&=(\beta_2+\beta_4x_2)(\Delta_{x_1}-\Delta_{x_2})\\
&=(\beta_2+\beta_4 x_2) \Bigg\{\big(1-\Delta^*_{x_2}\big)\Big[1-\sum_{k=1}^K\psi_k\exp{(-\lambda_{k,x_2})}\Big]-\big(1-\Delta^*_{x_1}\big)\Big[1-\sum_{k=1}^K\psi_k\exp{(-\lambda_{k,x_1})}\Big]\Bigg\},\\
\text{NDE}&=E\big(Y_{x_2 B_{x_1}M(x_1,B_{x_1})}-Y_{x_1 B_{x_1}M(x_1,B_{x_1})}\big) \\
&=\beta_3(x_2-x_1)+\beta_4(x_2-x_1)(1-\Delta_{x_1})+\beta_5(x_2-x_1)E(M(x_1,B_{x_1}))\\
&=(x_2-x_1)\Bigg\{\beta_3+\beta_4\big(1-\Delta^*_{x_1}\big)\Big(1-\sum_{k=1}^K\psi_k\exp{(-\lambda_{k,x_1})}\Big)+\beta_5(1-\Delta^*_{x_1})\sum_{k=1}^K\psi_k\lambda_{k,x_1}\Bigg\}\\
&=(x_2-x_1)\Bigg\{\beta_3+(1-\Delta^*_{x_1})\bigg[\beta_4\Big(1-\sum_{k=1}^K\psi_k\exp{(-\lambda_{k,x_1})}\Big)+\beta_5\sum_{k=1}^K\psi_k\lambda_{k,x_1}\bigg]\Bigg\} .
\end{align*}

\subsubsection{Log-likelihood function}
For group 1 consisting of subjects with positive, observed mediator values (i.e., $r_i=1$ and $m_i^*=m_i>0$), the log-likelihood contribution from the $i$th individual in component $k$ of the Poison mixture can be calculated as:
\begin{align*}
\ell_{ik}^1&=\log{\big(f(y_i,m_i^*,r_i=1|x_i,c_i=k)\big)}\\
&=\log(f(y_i, r_i=1|m_i^*,x_i,c_i=k)f(m_i^*|x_i,c_i=k))\\
&=\log(f(y_i|m_i^*,x_i,c_i=k)f(r_i=1|m_i^*,x_i,c_i=k)f(m_i^*|x_i,c_i=k))\\
&=\log(f(y_i|m_i^*,x_i))+\log(f(r_i=1|m_i^*))+\log(f(m_i^*|x_i,c_i=k))\\
&=\log(f(y_i|m_i^*,x_i))+\log(P(M_i^*>0|m_i^*))+\log(f(m_i^*|x_i,c_i=k))\\
&=-\log(\delta)-\frac{(y_i-\beta_0-\beta_1m_i^*-\beta_2-(\beta_3+\beta_4)x_i-\beta_5x_im_i^*)^2}{2\delta^2}+\log\Big[1-1_{(m_i^*\le L)}\exp{(-\eta^2m_i^*)}\Big]\\
&\hspace{0.4cm}+m_i^*\log{(\lambda_{ik})}-\log{(\exp{(\lambda_{ik})-1})}-\log{(m_i^*!)} - 0.5\log(2\pi).
\end{align*}
For group 2 consisting of subjects with observed mediator values as zeros (i.e., $r_i=0$ and $m_i^*=0$), the true mediator values could be true zeros or false zeros. The log-likelihood of the portion of individuals with a true zero mediator values ($m_i^*=m_i=0$) is
\begin{align*}
\ell_{i0}^2 &= \log(f(y_i,m_i^*=0,r_i=0|x_i,c_i=0))\\
&= \log(f(y_i|m_i^*=0,x_i,c_i=0)f(m_i^*=0,r_i=0|x_i,c_i=0))\\
&= \log(f(y_i|m_i^*=0,x_i)f(r_i=0|m_i^*=0,x_i,c_i=0)f(m_i^*=0|x_i,c_i=0))\\
&= \log(f(y_i|m_i^*=0,x_i)f(r_i=0|m_i^*=0)f(m_i^*=0|x_i,c_i=0))\\
&= \log(f(y_i|m_i^*=0,x_i)\cdot 1 \cdot 1)\\
&=-\log(\delta)-\frac{(y_i-\beta_0-\beta_3x_i)^2}{2\delta^2}-0.5\log{(2\pi)}.
\end{align*}
The rest of group 2 are individuals who had non-zero mediator values but were falsely observed as zeros ($m_i>0$ and $m_i^*=0$). To handle false zeros, a summation is required to account for all possibly observed, positive mediator values ranged from 1 to the constant L chosen in the mechanism of falsely observing zeros. Their log-likelihood contribution can be calculated as:
\begin{align*}
\ell_{ik}^2&=\log(f(y_i,m_i^*=0,r_i=0|x_i,c_i=k))\\
&=\log\bigg(\sum_{m=1}^{L} f(y_i, r_i=0|m,x_i,c_i=k)f(m|x_i,c_i=k)\bigg)\\
&=\log\bigg(\sum_{m=1}^{L} f(y_i|m,x_i,c_i=k) f(r_i=0|m,x_i,c_i=k)f(m|x_i,c_i=k)\bigg)\\
&=\log\bigg(\sum_{m=1}^{L} f(y_i|m,x_i) f(r_i=0|m)f(m|x_i,c_i=k)\bigg)\\
&=\log\Bigg\{\sum_{m=1}^{L} \frac{1}{\delta\sqrt{2\pi}}\exp{\bigg[-\frac{(y_i-\beta_0-\beta_1m-\beta_2-(\beta_3+\beta_4)x_i-\beta_5x_im)^2}{2\delta^2}\bigg]}\\
&\hspace{2.1cm}\cdot \exp{(-\eta^2m)}\frac{\lambda_{ik}^m}{m!(\exp{(\lambda_{ik})}-1)}\Bigg\}\\
&=\log\Bigg(\frac{1}{\delta\sqrt{2\pi}(\exp{(\lambda_{ik})}-1)}\sum_{m=1}^{L}h_{ik}(m)\Bigg)\\
&=-\log(\delta)-0.5\log(2\pi)-\log{(\exp{(\lambda_{ik})-1})}+\log\Bigg(\sum_{m=1}^{L}h_{ik}(m)\Bigg),\\
\end{align*}
where 
\begin{align*}
h_{ik}(m)=\frac{1}{m!}\exp\bigg(m\log{(\lambda_{ik})}-\frac{(y_i-\beta_0-\beta_1m-\beta_2-(\beta_3+\beta_4)x_i-\beta_5x_im)^2}{2\delta^2}-\eta^2m\bigg).
\end{align*}
The complete likelihood form in Equation (\ref{completell}) can be used. Again, the EM algorithm is used to obtain MLE of model parameters.

\subsubsection{Expectation step (E step)}
For group 1, let $\tau_{ik}^1(\Theta^0)=E_{c_i|y_i,m_i^*,r_i=1,x_i,\Theta^0}\big(1_{(c_i=k)}\big)$ for $k=1,\dots,K$. We have
\begin{align*}
\tau_{ik}^1(\Theta^0)&=E\big(1_{(c_i=k)}|y_i,m_i^*,r_i=1,x_i,\Theta^0\big)\\
&=P\big(c_i=k|y_i,m_i^*,r_i=1,x_i,\Theta^0\big)\\
&=\frac{f(y_i,m_i^*,r_i=1,c_i=k|x_i,\Theta^0)}{f(y_i,m_i^*=0,r_i=0|x_i,\Theta^0)}\\
&=\frac{f(y_i,m_i^*,r_i=1|c_i=k,x_i,\Theta^0)P(c_i=k|x_i,\Theta^0)}{\sum_{j=1}^Kf(y_i,m_i^*,r_i=1|c_i=j,x_i,\Theta^0)P(c_i=j|x_i,\Theta^0)}\\
&=\frac{f(y_i|m_i^*,x_i,\Theta^0)f(r_i=1|m_i^*,\Theta^0)f(m_i^*|c_i=k,x_i,\Theta^0)P(c_i=k|x_i,\Theta^0)}{\sum_{j=1}^Kf(y_i|m_i^*,x_i,\Theta^0)f(r_i=1|m_i^*,\Theta^0)f(m_i^*|c_i=j,x_i,\Theta^0)P(c_i=j|x_i,\Theta^0)}\\
&=\frac{f(m_i^*|c_i=k,x_i,\Theta^0)P(c_i=k|x_i,\Theta^0)}{\sum_{j=1}^Kf(m_i^*|c_i=j,x_i,\Theta^0)P(c_i=j|x_i,\Theta^0)}\\
&=\frac{\psi_k^0\frac{\lambda_{ik}^{m_i^*,0}}{m_i^*!\big(\exp{(\lambda_{ik}^0)-1\big)}}}{\sum_{j=1}^K\psi_j^0\frac{\lambda_{ij}^{m_i^*,0}}{m_i^*!\big(\exp{(\lambda_{ij}^0)-1\big)}}}\\
&=\frac{\psi_k^0\frac{\lambda_{ik}^{m_i^*,0}}{\big(\exp{(\lambda_{ik}^0)-1\big)}}}{\sum_{j=1}^K\psi_j^0\frac{\lambda_{ij}^{m_i^*,0}}{\big(\exp{(\lambda_{ij}^0)-1\big)}}}.
\end{align*}
For group 2, let $\tau_{ik}^2(\Theta^0)=E_{c_i|y_i,m_i^*=0,r_i=0,x_i,\Theta^0}\big(1_{(c_i=k)}\big)$ for $k=0,1,\dots,K$. We have
\begin{align*}
\tau_{ik}^2(\Theta^0)&=E\big(1_{(c_i=k)}|y_i,m_i^*=0,r_i=0,x_i,\Theta^0\big)\\
&=P\big(c_i=k|y_i,m_i^*=0,r_i=0,x_i,\Theta^0\big)\\
&=\frac{f(y_i,m_i^*=0,r_i=0,c_i=k|x_i,\Theta^0)}{f(y_i,m_i^*=0,r_i=0|x_i,\Theta^0)}\\
&=\frac{f(y_i,m_i^*=0,r_i=0|c_i=k,x_i,\Theta^0)P(c_i=k|x_i,\Theta^0)}{\sum_{j=0}^Kf(y_i,m_i^*=0,r_i=0|c_i=j,x_i,\Theta^0)P(c_i=j|x_i,\Theta^0)}\\
&=\frac{\Psi_{ik}\exp{(\ell_{ik}^2)}}{\sum_{j=0}^K\Psi_{ij}\exp{(\ell_{ij}^2)}}\Bigg|_{\hspace{0.1cm}\text{evaluated at}\hspace{0.1cm} \Theta^0}.
\end{align*}
Similarly, the M step can be performed as described in subsection \ref{mstep}.

\subsection{Model for Zero-inflated negative binomial mixture (ZINBM) mediators}
\subsubsection{Mediation effect and direct effect}
For ZINBM mediators, we also plug the equations (\ref{ymodel}) and (\ref{zinbm})-(\ref{transzinb2}) into definitions of average NIE and NDE for $X$ changing from $x_1$ to $x_2$ in equations (\ref{defNIE})-(\ref{defNDE}), 
\begin{align*}
\text{NIE}&=E\big(Y_{x_2B_{x_2}M(x_2,B_{x_2})}-Y_{x_2B_{x_1}M(x_1,B_{x_1})}\big) \\
&=E\big(Y_{x_2B_{x_2}M(x_2,B_{x_2})} - Y_{x_2B_{x_2}M(x_1,B_{x_1})}\big) + E\big(Y_{x_2B_{x_2}M(x_1,B_{x_1})}  - Y_{x_2B_{x_1}M(x_1,B_{x_1})}\big) \\
&=\text{NIE}_1+\text{NIE}_2,\\
\text{NIE}_1&=E\big(Y_{x_2B_{x_2}M(x_2,B_{x_2})} - Y_{x_2B_{x_2}M(x_1,B_{x_1})}\big) \\
&=(\beta_1+\beta_5x_2)\Big(E(M(x_2,B_{x_2}))-E(M(x_1,B_{x_1}))\Big)\\
&=(\beta_1+\beta_5x_2)\Bigg[\sum_{m=0}^\infty mf(m;x_2)-\sum_{m=0}^\infty mf(m;x_1)\Bigg]\\
&=(\beta_1+\beta_5x_2) \Bigg[(1-\Delta^*_{x_2})\sum_{k=1}^K\psi_k\mu_{x_2,k}-(1-\Delta^*_{x_1})\sum_{k=1}^K\psi_k\mu_{x_1,k}\Bigg],\\
\text{NIE}_2&=E\big(Y_{x_2B_{x_2}M(x_1,B_{x_1})}  - Y_{x_2B_{x_1}M(x_1,B_{x_1})}\big)\\
&=(\beta_2+\beta_4x_2)(E(B_{x_2})-E(B_{x_1}))\\
&=(\beta_2+\beta_4x_2)(\Delta_{x_1}-\Delta_{x_2})\\
&=(\beta_2+\beta_4 x_2) \Bigg\{\big(1-\Delta^*_{x_2}\big)\Big[1-\sum_{k=1}^K\psi_k(\frac{r}{r+\mu_{x_2,k}})^{r}\Big]-\big(1-\Delta^*_{x_1}\big)\Big[1-\sum_{k=1}^K\psi_k(\frac{r}{r+\mu_{x_1,k}})^{r}\Big]\Bigg\},\\
\text{NDE}&=E\big(Y_{x_2 B_{x_1}M(x_1,B_{x_1})}-Y_{x_1 B_{x_1}M(x_1,B_{x_1})}\big) \\
&=\beta_3(x_2-x_1)+\beta_4(x_2-x_1)(1-\Delta_{x_1})+\beta_5(x_2-x_1)E(M(x_1,B_{x_1}))\\
&=(x_2-x_1)\Bigg\{\beta_3+\beta_4\big(1-\Delta^*_{x_1}\big)\Big(1-\sum_{k=1}^K\psi_k(\frac{r}{r+\mu_{x_1,k}})^{r}\Big)+\beta_5(1-\Delta^*_{x_1})\sum_{k=1}^K\psi_k\mu_{x_1,k}\Bigg\}\\
&=(x_2-x_1)\Bigg\{\beta_3+(1-\Delta^*_{x_1})\bigg[\beta_4\Big(1-\sum_{k=1}^K\psi_k(\frac{r}{r+\mu_{x_1,k}})^{r}\Big)+\beta_5\sum_{k=1}^K\psi_k\mu_{x_1,k}\bigg]\Bigg\}.
\end{align*}

\subsubsection{Log-likelihood function}
The log-likelihood contribution from the $i$th individual from NB mixture component $k$ in group 1 (i.e. $r_i=1$, $m_i=m_i^*>0$) can be calculated as:
\begin{align*}
\ell_{ik}^1&=\log{\big(f(y_i,m_i^*,r_i=1|x_i,c_i=k)\big)}\\
&=\log(f(y_i, r_i=1|m_i^*,x_i,c_i=k)f(m_i^*|x_i,c_i=k))\\
&=\log(f(y_i|m_i^*,x_i,c_i=k)f(r_i=1|m_i^*,x_i,c_i=k)f(m_i^*|x_i,c_i=k))\\
&=\log(f(y_i|m_i^*,x_i))+\log(f(r_i=1|m_i^*))+\log(f(m_i^*|x_i,c_i=k))\\
&=\log(f(y_i|m_i^*,x_i))+\log(P(M_i^*>0|m_i^*))+\log(f(m_i^*|x_i,c_i=k))\\
&=-\log(\delta)-\frac{(y_i-\beta_0-\beta_1m_i^*-\beta_2-(\beta_3+\beta_4)x_i-\beta_5x_im_i^*)^2}{2\delta^2}-0.5\log(2\pi)\\
&\hspace{0.4cm}+\log\Big[1-1_{(m_i^*\le L)}\exp{(-\eta^2m_i^*)}\Big]+\log{(\Gamma(r+m_i^*))}-\log{(\Gamma(r))}-\log{(m_i^*!)}\\
&\hspace{0.4cm}+m_i^*\log\bigg(\frac{\mu_{ik}}{r+\mu_{ik}}\bigg)+r\log\bigg(\frac{r}{r+\mu_{ik}}\bigg)-\log\Bigg[1-\bigg(\frac{r}{r+\mu_{ik}}\bigg)^{r}\Bigg].
\end{align*}

For group 2, the log-likelihood of the portion of individuals with a true zero mediator values (i.e., $r_i=0$, $m_i=m_i^*=0$) is
\begin{align*}
\ell_{i0}^2 &= \log(f(y_i,m_i^*=0,r_i=0|x_i,c_i=0))\\
&= \log(f(y_i|m_i^*=0,x_i,c_i=0)f(m_i^*=0,r_i=0|x_i,c_i=0))\\
&= \log(f(y_i|m_i^*=0,x_i)f(r_i=0|m_i^*=0,x_i,c_i=0)f(m_i^*=0|x_i,c_i=0))\\
&= \log(f(y_i|m_i^*=0,x_i)f(r_i=0|m_i^*=0)f(m_i^*=0|x_i,c_i=0))\\
&= \log(f(y_i|m_i^*=0,x_i)\cdot 1 \cdot 1)\\
&=-\log(\delta)-\frac{(y_i-\beta_0-\beta_3x_i)^2}{2\delta^2}-0.5\log{(2\pi)}.
\end{align*}
The rest of group 2 are individuals who had non-zero mediator values (from NB mixture) but were falsely observed as zeros (i.e., $r_i=0$, $m_i>0$, and $m_i^*=0$). Given the unknown true values of false zeros, a summation is used to account for possible non-zero mediator values that are less than or equal to the constant $L$ chosen in the mechanism of falsely observing zeros. The log-likelihood contribution from $k$th mixture component in group 2 can be calculated as:
\begin{align*}
\ell_{ik}^2&=\log(f(y_i,m_i^*=0,r_i=0|x_i,c_i=k))\\
&=\log\bigg(\sum_{m=1}^{L} f(y_i, r_i=0|m,x_i,c_i=k)f(m|x_i,c_i=k)\bigg)\\
&=\log\bigg(\sum_{m=1}^{L} f(y_i|m,x_i,c_i=k) f(r_i=0|m,x_i,c_i=k)f(m|x_i,c_i=k)\bigg)\\
&=\log\bigg(\sum_{m=1}^{L} f(y_i|m,x_i) f(r_i=0|m)f(m|x_i,c_i=k)\bigg)\\
&=\log\Bigg\{\sum_{m=1}^{L} \frac{1}{\delta\sqrt{2\pi}}\exp{\bigg[-\frac{(y_i-\beta_0-\beta_1m-\beta_2-(\beta_3+\beta_4)x_i-\beta_5x_im)^2}{2\delta^2}\bigg]}\\
&\hspace{2.1cm}\cdot \exp{(-\eta^2m)} \frac{\Gamma(r+m)}{\Gamma(r)m!}\Big(\frac{\mu_{ik}}{r+\mu_{ik}}\Big)^m\frac{1}{(\frac{r}{r+\mu_{ik}})^{-r}-1}\Bigg\}\\
&=\log\Bigg(\frac{1}{\delta\sqrt{2\pi}\Big[\big(\frac{r}{r+\mu_{ik}}\big)^{-r}-1\Big]}\sum_{m=1}^{L}h_{ik}(m)\Bigg)\\
&=-\log(\delta)-0.5\log(2\pi)-\log{\Big[\big(\frac{r}{r+\mu_{ik}}\big)^{-r}-1\Big]}+\log\Bigg(\sum_{m=1}^{L}h_{ik}(m)\Bigg),
\end{align*}
where 
\begin{align*}
h_{ik}(m)=\frac{\Gamma(r+m)}{\Gamma(r)m!}\Big(\frac{\mu_{ik}}{r+\mu_{ik}}\Big)^m\exp\bigg(-\frac{(y_i-\beta_0-\beta_1m-\beta_2-(\beta_3+\beta_4)x_i-\beta_5x_im)^2}{2\delta^2}-\eta^2m\bigg).
\end{align*}
The complete likelihood form in Equation (\ref{completell}) can be used. Again, the EM algorithm is used to obtain MLE of model parameters.

\subsubsection{Expectation step (E step)}
For group 1, let $\tau_{ik}^1(\Theta^0)=E_{c_i|y_i,m_i^*,r_i=1,x_i,\Theta^0}\big(1_{(c_i=k)}\big)$ for $k=1,\dots,K$. We have
\begin{align*}
\tau_{ik}^1(\Theta^0)&=E\big(1_{(c_i=k)}|y_i,m_i^*,r_i=1,x_i,\Theta^0\big)\\
&=P\big(c_i=k|y_i,m_i^*,r_i=1,x_i,\Theta^0\big)\\
&=\frac{f(y_i,m_i^*,r_i=1,c_i=k|x_i,\Theta^0)}{f(y_i,m_i^*=0,r_i=0|x_i,\Theta^0)}\\
&=\frac{f(y_i,m_i^*,r_i=1|c_i=k,x_i,\Theta^0)P(c_i=k|x_i,\Theta^0)}{\sum_{j=1}^Kf(y_i,m_i^*,r_i=1|c_i=j,x_i,\Theta^0)P(c_i=j|x_i,\Theta^0)}\\
&=\frac{f(y_i|m_i^*,x_i,\Theta^0)f(r_i=1|m_i^*,\Theta^0)f(m_i^*|c_i=k,x_i,\Theta^0)P(c_i=k|x_i,\Theta^0)}{\sum_{j=1}^Kf(y_i|m_i^*,x_i,\Theta^0)f(r_i=1|m_i^*,\Theta^0)f(m_i^*|c_i=j,x_i,\Theta^0)P(c_i=j|x_i,\Theta^0)}\\
&=\frac{f(m_i^*|c_i=k,x_i,\Theta^0)P(c_i=k|x_i,\Theta^0)}{\sum_{j=1}^Kf(m_i^*|c_i=j,x_i,\Theta^0)P(c_i=j|x_i,\Theta^0)}\\
&=\psi_k\frac{\frac{\Gamma(r+m_i^*)}{\Gamma(r)m_i^*!}(\frac{\mu_{ik}}{r+\mu_{ik}})^{m_i^*}}{(\frac{r}{r+\mu_{ik}})^{-r}-1}\Bigg/\sum_{j=1}^K\psi_j\frac{\frac{\Gamma(r_j+m_i^*)}{\Gamma(r_j)m_i^*!}(\frac{\mu_{ij}}{r+\mu_{ij}})^{m_i^*}}{(\frac{r_j}{r_j+\mu_{ij}})^{-r_j}-1}\Bigg|_{\hspace{0.1cm}\text{evaluated at}\hspace{0.1cm} \Theta^0}.
\end{align*}
For group 2, let $\tau_{ik}^2(\Theta^0)=E_{c_i|y_i,m_i^*=0,r_i=0,x_i,\Theta^0}\big(1_{(c_i=k)}\big)$ for $k=0,1,\dots,K$. We have
\begin{align*}
\tau_{ik}^2(\Theta^0)&=E\big(1_{(c_i=k)}|y_i,m_i^*=0,r_i=0,x_i,\Theta^0\big)\\
&=P\big(c_i=k|y_i,m_i^*=0,r_i=0,x_i,\Theta^0\big)\\
&=\frac{f(y_i,m_i^*=0,r_i=0,c_i=k|x_i,\Theta^0)}{f(y_i,m_i^*=0,r_i=0|x_i,\Theta^0)}\\
&=\frac{f(y_i,m_i^*=0,r_i=0|c_i=k,x_i,\Theta^0)P(c_i=k|x_i,\Theta^0)}{\sum_{j=0}^Kf(y_i,m_i^*=0,r_i=0|c_i=j,x_i,\Theta^0)P(c_i=j|x_i,\Theta^0)}\\
&=\frac{\Psi_{ik}\exp{(\ell_{ik}^2)}}{\sum_{j=0}^K\Psi_{ij}\exp{(\ell_{ij}^2)}}\Bigg|_{\hspace{0.1cm}\text{evaluated at}\hspace{0.1cm} \Theta^0}.
\end{align*}
Finally we get 
\begin{align}
Q(\Theta|\Theta^0)&=E(\ell|y_i,m_i^*,r_i,x_i,\Theta^0)\nonumber\\
&=\sum_{i\in\text{group1}}\sum_{k=1}^K\tau_{ik}^1(\Theta^0)\Big(\log{(\Psi_{ik})}+\ell_{ik}^1\Big)+\sum_{i\in\text{group2}}\sum_{k=0}^K\tau_{ik}^2(\Theta^0)\Big(\log{(\Psi_{ik})}+\ell_{ik}^2\Big).
\end{align}
Similarly, the M step can be performed as described in subsection \ref{mstep}.

\bibliography{draft_ref_mix}

\end{document}